\documentclass{aa}

\usepackage[applemac]{inputenc} % accents sous mac

\usepackage{subfig}
\usepackage{natbib}  %pour la bibliographie
\bibpunct{(}{)}{;}{a}{}{,} % to follow the A&A style
\usepackage{array} %options pour les tableaux
\usepackage{tabularx}
\usepackage{multirow}
\usepackage{graphicx}
\usepackage[version=3]{mhchem}
\usepackage{txfonts}
\usepackage{amsmath}
\usepackage{lscape}

\newcommand{\tablenotea}[1]{\parbox{8.5cm}{ \indent
\footnotesize{\textsc{Note.--}~#1}}}

\begin{document}

\title{A chemical model for the atmosphere of hot Jupiters}
\author{O. Venot\inst{\ref{LAB},} \inst{\ref{CNRS}}
\and E. H\'{e}brard\inst{\ref{LAB},}\inst{\ref{CNRS}}
\and M. Ag\'{u}ndez\inst{\ref{LAB},}\inst{\ref{CNRS}}
\and M. Dobrijevic\inst{\ref{LAB},}\inst{\ref{CNRS}}
\and F. Selsis\inst{\ref{LAB},}\inst{\ref{CNRS}}
\and F. Hersant\inst{\ref{LAB},}\inst{\ref{CNRS}}
\and N. Iro\inst{\ref{keele},}\inst{\ref{LESIA}}
\and R. Bounaceur\inst{\ref{LRGP}}}

\institute{Univ. Bordeaux, LAB, UMR 5804, F-33270, Floirac, France\\\email{venot@obs.u-bordeaux1.fr}\label{LAB}
\and CNRS, LAB, UMR 5804, F-33270, Floirac, France\label{CNRS}
\and Astrophysics Group, Keele University, Keele, Staffordshire, ST5 5BG, UK\label{keele}
\and LESIA, Observatoire de Paris, 5 Place Jules Janssen, F-92195 Meudon, France\label{LESIA}
\and Laboratoire R\'{e}actions et G\'{e}nie des Proc\'{e}d\'{e}s, LRGP UPR 3349 CNRS, Universit\'{e} de Lorraine, 1 rue Grandville, BP 20401, F-54001 Nancy, France\label{LRGP}}

\date{Received <date> /
Accepted <date>}

\abstract{%Context
The atmosphere of hot Jupiters can be probed by primary transit and secondary eclipse spectroscopy. Due to the intense UV irradiation, mixing and circulation, their chemical composition is maintained out of equilibrium and must be modeled with kinetic models.}
{%Aims
Our purpose is to release a chemical network, and the associated rate coefficients, developed for the temperature and pressure range relevant to hot Jupiters atmospheres. Using this network, we study the vertical atmospheric composition of the two hot Jupiters (HD~209458b and HD~189733b) with a model that includes photolyses and vertical mixing and we produce synthetic spectra.}
{%Methods
The chemical scheme is derived from applied combustion models that have been methodically validated over a range of temperatures and pressures typical of the atmospheric layers influencing the observations of hot Jupiters. We compare the predictions obtained from this scheme with equilibrium calculations, with different schemes available in the literature that contain N-bearing species and with previously published photochemical models.}
{%Results
Compared to other chemical schemes that were not subjected to the same systematic validation, we find significant differences whenever non-equilibrium processes take place (photodissociations or vertical mixing). The deviations from the equilibrium, and thus the sensitivity to the network, are more important for HD~189733b, as we assume a cooler atmosphere than for HD~209458b. We found that the abundances of \ce{NH3} and HCN can vary by two orders of magnitude depending on the network, demonstrating the importance of comprehensive experimental validation. A spectral feature of \ce{NH3} at 10.5~$\mu$m is sensitive to these abundance variations and thus to the chemical scheme.}
{%Conclusions
Due to the influence of the kinetics, we recommend the use of a validated scheme to model the chemistry of exoplanet atmospheres. The network we release is robust for temperatures within 300-2500~K and pressures from 10 mbar up to a few hundreds of bars, for species made of C, H, O, N. It is validated for species up to 2 carbon atoms and for the main nitrogen species (\ce{NH3}, HCN, \ce{N2}, NO$_X$). 
Although the influence of the kinetic scheme on the hot Jupiters spectra remains within the current observational error bars (with the exception of \ce{NH3}), it will become more important for atmospheres that are cooler or subjected to higher UV fluxes, departing more from equilibrium.}


\maketitle

\section{Introduction}

So far, more than 700 exoplanets have been confirmed and thousands of transiting candidates have been identified by the space telescope Kepler \citep{2012arXiv1202.5852B}. Among them, hot Jupiters are a class of gas giants with orbital periods of a few days or less. They are found around $\sim 0.5\%$ of KGF stars \citep{howard2010occurrence, howard2011}. About $10\%$ of them transit their host star and their atmospheric composition and physical structure can be studied by transit spectroscopy \citep[e.g.][]{charbonneau2000detection, charbonneau2008broadband, richardson2007spectrum, tinetti2007water, sing2008hubble, swain2008mid, swain2008presence, swain2009water, swain2009molecular, 2012arXiv1202.4721H}. 

Although current observations are still limited and subjected to divergent interpretations, future instruments such as E-ELT, JWST \citep{Gardner2006}, EChO \citep{tinetti-Echo}, FINESSE \citep{swain2010finesse} should provide better constraints on both the chemical composition and the temperature profiles of the nearby hot Jupiters like HD~189733b and HD~209458b. They will also be able to study more distant targets and deliver statistically significant trends about the nature of their atmospheres. Chemical modeling will be an important component of these studies. It will point to key observations able to distinguish between various hypotheses and will be used to analyze the observations and constrain, for instance, the atmospheric elemental abundances.

\begin{figure}[!htbp]
\centering
{\includegraphics[width=0.95\columnwidth]{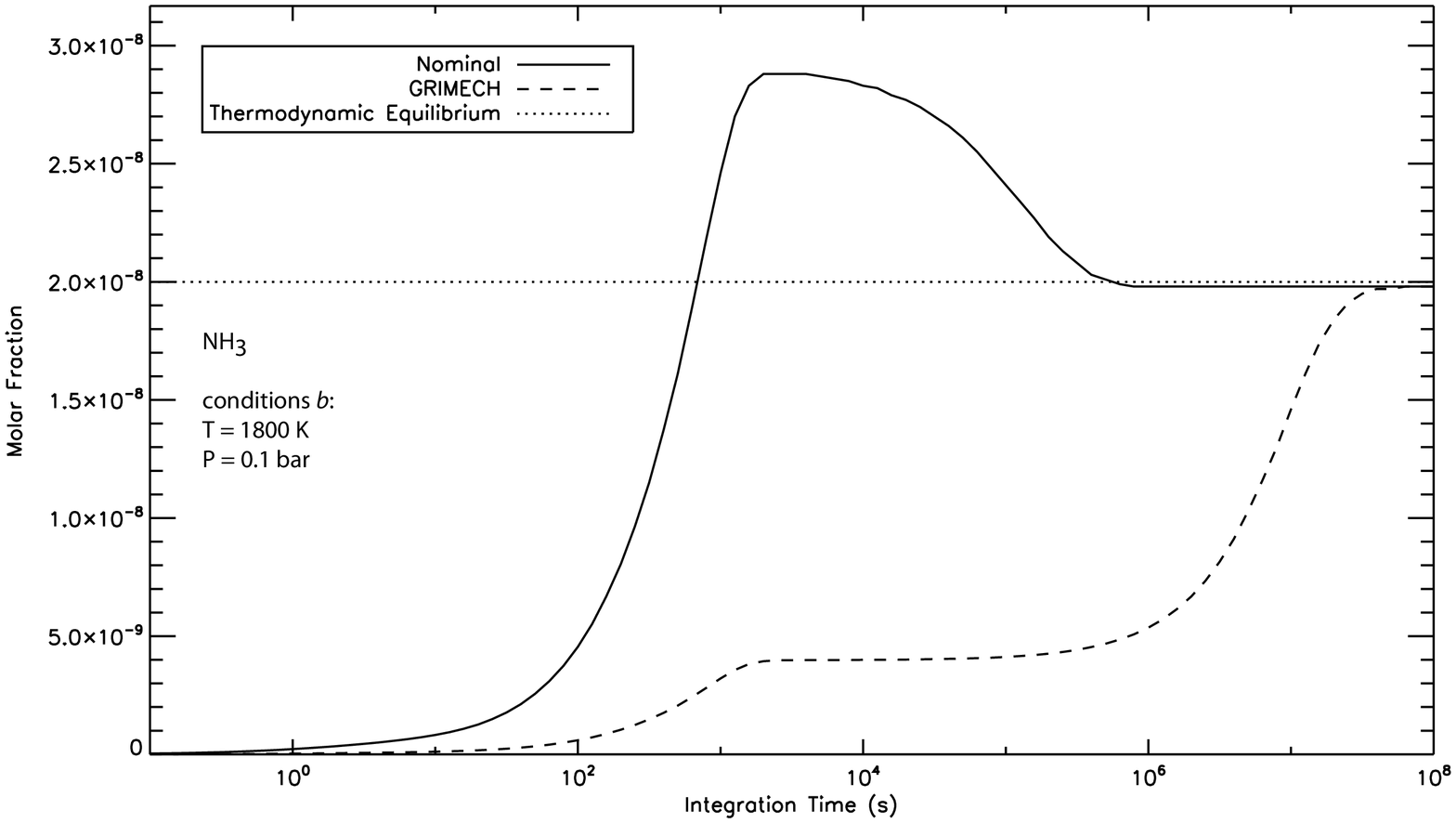}}
\caption{Abundances of \ce{NH3} as a function of time computed with two kinetic schemes fully reversed according to equilibrium constants but differing by their nitrogen chemistry (nominal and \textit{GRIMECH} as defined in Sects.~\ref{sec:nitrogen} and \ref{sec:nitrogen2}). While they both converge towards the equilibrium (dotted line) they exhibit very different evolution. Initial condition is a mixture of \ce{H2}, \ce{CH4}, \ce{O2}, \ce{N2} and \ce{He} with solar elemental abundances.}
\label{fig:0D_GRIM_NOM}
\end{figure}

The first models of hot Jupiter atmospheres assumed chemical equilibrium \citep[e.g.][]{burrows1999chemical, seager2000theoretical, sharp2007atomic, 2007ApJ...661L.191B, burrows2007theoretical, burrows2008theoretical, fortney2008unified}. However, strongly irradiated atmospheres are unlikely to be at chemical equilibrium. Their intense UV irradiation (typically 10,000 times the flux received on the top of the atmosphere of Jupiter) and strong dynamics result in photolyses and diffusion/advection timescales that are comparable or shorter than the chemical ones.  Deviations from the thermodynamic equilibrium have been discussed with timescale arguments \citep[e.g.][]{lodders2002atmospheric, fortney2006influence, fortney2008synthetic, visscher2006, visscher2010, madhusudhan2010inversion}, or modeled with a few reactions describing the CO-\ce{CH4} conversion coupled with the dynamics \citep{cooper2006dynamics}. A more detailed modeling requires the use of a photochemical kinetic network. A kinetic network is, in practice, a list of reactions and associated rate coefficients able to describe quantitatively (within a certain accuracy) the kinetics of a pool of species, usually the most abundant ones. Constructing such network of reactions implies to solve two major questions. One has to do with the completeness of the network: What are the species and the reactions connecting them that must be included? The other issue is the availability of the kinetic data: The literature and databases may not provide the rate coefficients for some of the needed reactions or may provide conflicting values with no recommendation. These two issues are tightly connected and both depend on the considered range of temperatures and pressures. Eventually, and whatever the methodology adopted to select the reactions and their rates, it is the ability to predict experimentaly-controled abundances that can validate or not the network.\\
To investigate the consequences of the strong UV incident flux on neutral species, photochemical models have then been developed \citep{liang2003source, liang2004insignificance}. Based on kinetics model dedicated to Jupiter's low-temperature atmosphere, these models however neglected endothermic reactions, which are in fact rather efficient in such hot atmospheres. \citet{line2010high} introduced some endothermic reactions to a similar Jovian photochemical scheme but most of the pre-existing reactions were not reversed. They were therefore not able to reproduce the thermodynamic equilibrium, which occurs in the deep atmospheres of hot Jupiters. \citet{zahnle2009atmospheric, zahnle2009soots}  developed a photochemical model considering the reversal of their whole set of two-bodies exothermic reactions. They selected their rate constants in the NIST database\footnote{http://www.kinetics.nist.gov/kinetics} based on the following criteria: relevance of temperature conditions, date of review, date of the experiment and date of the theoretical study (in order of preference). \citet{moses2011disequilibrium} developed a model that considered the reversal of all the reactions, including three-body reactions, ensuring to reproduce the thermodynamical equilibrium from the top to the deepest layers of the atmosphere. Their chemical scheme is derived from the Jupiter and Saturn models \citep{gladstone1996hydrocarbon, moses1996, moses1995post, moses1995nitrogen, moses2000photochemistry, moses2000photochemistry2} with further updates on the basis of combustion-chemistry literature. None of the aforementioned works discuss the validation of the chemical scheme against experiments. In addition, the fact that computed abundances evolve towards the composition predicted by equilibrium calculations (at given pressure \textit{P} and temperature \textit{T} with no external irradiation nor mixing) is by no means a validation of the kinetic network. Indeed, any network containing at least as many independent reversible reactions as modeled species, in which the rates for the backward processes are derived from equilibrium constants and forward rates, will evolve toward the equilibrium predicted with the same equilibrium constants, whatever the quantitative values of the forward rates, as illustrated in Fig.~\ref{fig:0D_GRIM_NOM}.
Fortunately, and due to the physical conditions and elemental composition of hot Jupiters (and hot Neptunes) atmospheres, we benefit from decades of intensive work done in the field of combustion, that includes a vast amount of experiments, the development of comprehensive mechanisms\footnote{In the field of combustion, a \textit{mechanism} or {reaction base} is a network of reactions able to describe the kinetic evolution of a given pool of species. The mechanism includes the list of reactions and the associated rate coefficients, in a modified Arrhenius form, as well as the thermodynamic data for all the species involved in these reactions, which are required to calculate the equilibrium constants of the reactions and the rates of the reverse reactions} and the systematic comparison between the two.
Therefore, we propose in the present work a new mechanism dedicated to the chemical modeling of hot atmospheres that is not adapted from previous Solar System photochemical models but derives from industrial applications (mainly combustion in car engines). Details about this chemical network and its range of validity are presented in Sect.~\ref{sec:kinetic}.

We use this chemical network in a 1D model that includes photolyses and vertical transport, which has been previously used to study the atmospheric photochemistry of various objects of the Solar System: Neptune \citep{dobrijevic2010key}, Titan \citep{hebrard2006photochemical,hebrard2007photochemical}, Saturn \citep{dobrijevic2003effect, cavalie2009first}, and Jupiter \citep{cavalie2008observation}) as well as extrasolar terrestrial planets \citep{selsis2002signature}.
We model the photochemistry of two hot Jupiters: HD~209458b and HD~189733b (Sect.~\ref{Sec:application}). We study the departure from thermodynamic equilibrium and compare our results with those of \citet{moses2011disequilibrium} (Sect.~\ref{sec:results}). We also investigate how including different reaction networks specific to nitrogen-bearing species influences the model results (Sect.~\ref{sec:nitrogen2}) and the planetary synthetic spectra (Sect.~\ref{sec:spectra}).

\section{The model}\label{sec:model}
 
\subsection{Kinetic network: from car engine to hot Jupiters}\label{sec:kinetic}

Significant progress has been done during the past decade in the development of validated combustion mechanisms. 
In the context of limiting the environmental impact of transportation, there is indeed  a need in the development of detailed chemical kinetic models more predictive and more accurate  for the combustion of fuels. 
One part of the studies undertaken in the LRGP (Laboratoire R\'{e}actions et G\'{e}nie des Proc\'{e}d\'{e}s, Nancy, France) concerns engine-fuel adaptation in order to improve the efficiency of engines and to limit the emission of pollutants. Gasoline and Diesel fuels contain a large number of molecules belonging to several major hydrocarbon families. Biofuels contain also oxygenated species, such as alcohols and methyl esters. Oxidation and combustion of these complex blends occur by radical chain reactions involving hundreds of molecular and radical species and several thousands of elementary reactions in the case of pure reference fuels such as n-heptane, iso-octane or cetane. The primary focus of the currently developed chemical models is to simulate the main combustion parameters (auto-ignition delay times, laminar flame speed, heat release), which are needed for the design of engines or burners, to estimate the fuel consumption, and to model the formation of some of the main regulated pollutants (carbon monoxide, nitrogen oxides, unburned hydrocarbons and particulate matter).
Most of these kinetic models were developed for industrial applications and have been validated in a range of temperatures, from 300 to approximately 2500 K, and for pressures from 0.01 bar to some hundreds of bar. What is worth noticing is the similarity of these temperature and pressure ranges with the conditions prevailing in hot Jupiters atmospheres, in the very layers where they influence the observed molecular features. In addition, combustion mechanisms mainly deal with molecules made of C, H, O and N, which are also the main constituents of the molecules and radicals in these atmospheres.
For this reason, we have decided to implement such a mechanism, which has already been applied successfully to many cases and systematically validated  \citep{bounaceur2007kinetic}, to study the atmosphere of hot Jupiters.

In this study we have used a C/H/O/N mechanism, whose core is a C$_0$-C$_2$ mechanism that includes all the reactions required to model the kinetic evolution of radicals and molecules containing less than three carbon atoms. This mechanism also contains some species with more carbon atoms that are needed to model the abundance of C$_0$-C$_2$ species. This mechanism does not include nitrogen species, except \ce{N2} as a third body.
As nitrogen species, such as \ce{N2}, \ce{NH3}, \ce{HCN}, \ce{CN}, are expected to be important constituents of hot Jupiter atmospheres, we completed this C$_0$-C$_2$ base with a validated sub-mechanism specifically constructed to model nitrogen species and all the cross-term reactions involved (for instance, reactions between alkanes and NO$_{X}$). 
These mechanisms do not use rate coefficients that have been adjusted by optimization procedures in order to fit experiments. Their values are those recommended for the individual processes by the main kinetics databases for combustion \citep{tsang1986chemical, NIST, smithgri, baulch2005evaluated}. The list of the reactions and their rate coefficients are available in the online database KIDA: KInetic Database for Astrochemistry\footnote{http://kida.obs.u-bordeaux1.fr} \citep{wakelam_kida2012}. The final mechanism includes 957 reversible and 6 irreversible reactions (see Sect.~\ref{sec:reverse}), involving 105 neutral species (molecule or radical). Helium is also included in this mechanism and plays the role of third body in some reactions.

\subsubsection{C$_0$-C$_2$ reaction base}\label{sec:C0-C2}

The C$_0$-C$_2$ reaction base we use was developed for industrial applications and was first presented by \citet{barbe1995experimental} and has been continuously updated \citep{fournet1999experimental, bounaceur2010modeling}. This mechanism is designed to reproduce the kinetics of species with less than three carbons. It includes all the unimolecular or bimolecular reactions involving radicals or molecules containing no more than two carbon atoms. This mechanism has been built by using a reaction grid, as proposed by \citet{tsang1986chemical}. Every unimolecular and bimolecular elementary forward reactions involving the considered reacting species have been systematically written. Reacting species include 46 compounds (19 molecules and 27 radicals) which have been ranked according to the molecular formula O$_x$C$_y$H$_z$ (with $x$ varying from 0 to 3, $y$ from 0 to 2 and $z$ from 0 to 6): CO, \ce{H2}, \ce{H2O}, \ce{O2}, \ce{H2O2}, \ce{CH4}, \ce{H2CO}, \ce{CH3OH}, \ce{CO2}, \ce{CH3OOH}, \ce{C2H2}, \ce{C2H4}, \ce{C2H6}, \ce{CH2CO}, \ce{CH3CHO}, \ce{C2H5OH}, \ce{C2H5OOH}, \ce{CH3COOOH}, c\ce{C2H4O} (Ethylene Oxide), C, CH, \ce{^1CH2} (singlet),\ce{^3CH2} (triplet), O($^3$P), H, OH, OOH, \ce{CH3}, HCO, \ce{CH2OH}, \ce{CH3O}, \ce{CH3OO}, \ce{C2H}, \ce{C2H3}, \ce{C2H5}, CHCO, \ce{CH2CHO}, \ce{CH3CO}, \ce{C2H5O}, \ce{C2H4OOH}, \ce{C2H5OO}, \ce{CH3COOO}, \ce{CH3OCO}, \ce{CO2H}, 1-\ce{C2H4OH} and 2-\ce{C2H4OH} (ethyl radical isomers, 1-hydroxy and 2-hydroxy). This ranking permits to separate easily the part of the mechanism related to pyrolysis reactions from the one related to oxidation or combustion.
The mechanism also includes 14 species containing 3 or 4 carbon atoms: \ce{C3H8}, \ce{C4H8}, \ce{C4H10}, \ce{C2H5CHO}, \ce{C3H7OH}, \ce{C3H7O}, \ce{C4H9O}, \ce{C2H6CO}, \ce{C3H8CO}, \ce{C2H3CHO}, n-\ce{C3H7}, i-\ce{C3H7} (isopropyl and n-propyl radical isomers), 1-\ce{C4H9} and 2-\ce{C4H9} (1-butyl and 2-butyl radical isomers).

This C$_0$-C$_2$ mechanism has been widely validated in the 300-2500~K, 0.01-100~bar range for several types of reactors such as shock-tubes, perfectly stirred reactors, plug-flow reactors, rapid compression machines, laminar flames (e.g. \citealt{battin2006experimental, bounaceur2007kinetic, anderlohr2010thermal, bounaceur2010modeling, wang2010experimental}). Obviously, it is not possible to describe in details all these validations, but we can mention, for instance, the very recent work of \citet{dirrenbergermeasurements} who has studied experimentally and modeled with success the laminar burning velocity of several mixtures including air, hydrogen and components of natural gas. Laminar burning velocities are important parameters in many areas of combustion science such as the design of burners and the prediction of explosions. They also play an essential role in determining several important aspects of the combustion process in spark ignition engines. These experiments have been done in specific mixtures, containing nitrogen in the sole form of \ce{N2} and in which nitrogen species  produced from \ce{N2} (NO$_{X}$ in the typical mixtures used in combustion) do not significantly affect the outcome, in order to validate the C$_0$-C$_2$ mechanism itself. Therefore, a model including only the C$_0$-C$_2$ base would not be accurate to predict the abundance of  C$_0$-C$_2$ species in this range of $P$ and $T$ when applied to mixtures containing or producing (by reaction with \ce{N2}) significant levels of nitrogen species other than \ce{N2}.

\subsubsection{Nitrogen reaction base}\label{sec:nitrogen}

In our nominal model, the sub-network for the nitrogen bearing species is derived from \citet{konnov2000development, konnov2009implementation} and \citet{coppens2007effects}. It is based on a comprehensive analysis of the combustion chemistry of nitrogen oxides \citep{konnov1999kinetic}, ammonia \citep{konnov2000kinetic}, hydrazine \citep{konnov2001kinetic}, and modeling of nitrogen oxides formation in different combustion systems \citep{konnov1999, konnov2000b, konnov2001b}.
The mechanism was tested at the California Institute of Technology, USA, and found suitable for steady one-dimensional detonation and constant volume explosion simulations \citep{schultz1999}. It was also preferred by the researchers at the University of Bielefeld, Germany to analyze flame structure and NO reburning in C$_3$ flames \citep{atakan2000laser}. In addition, we consider a few additional pathways for HCN oxidation from \citet{dagaut2008oxidation}. 

Validations of our nominal sub-network for nitrogen bearing species have been made on the basis of experimental data obtained, for instance, by oxidation of HCN in a silica jet-stirred reactor (JSR) at atmospheric pressure and from 1000 to 1400 K \citep{dagaut2008oxidation}, or studying laminar flame speeds in \ce{NH3} - \ce{N2O} mixtures \citep{Brown19941011}. 
The nitrogen mechanism includes 42 species (molecule or radical): \ce{NO3}, \ce{HONO2}, \ce{CH3ONO}, \ce{CH3NO2}, \ce{HNO2}, \ce{CH3NO}, \ce{NO2}, HONO, HCNN, HCNO, \ce{N2O}, NCO, HNO, HOCN, NNH, \ce{H2CN}, N($^4$S), CN, HNCO, NO, NH, \ce{NH2}, HCN, \ce{NH3}, \ce{N2}, \ce{N2O4}, \ce{N2O3}, \ce{N2H2}, \ce{N2H3}, \ce{N2H4}, HNNO, HNOH, \ce{HNO3}, \ce{NH2OH}, \ce{H2NO}, CNN, \ce{H2CNO}, \ce{C2N2}, HCNH, HNC, HON and NCN.

For comparison, we have also used other nitrogen sub-mechanisms, which are presented in Sec. \ref{sec:nitrogen2} with the corresponding results.

Because the mechanism we use was created from individual processes and validated without any optimization of their reaction coefficients, its application outside the condition range of validation is not problematic. This is an issue, for instance, with the well-known combustion mechanism GRI-Mech 3.0\footnote{http://www.me.berkeley.edu/gri\_mech/} \citep{smithgri}, proposed by Gas Research Institute, which is an optimized mechanism designed to model natural gas combustion. Optimization makes the model extremely accurate within the optimization domain but its application beyond is risky \citep{battin2011towards}.

\subsubsection{Reversible reactions: kinetics vs thermodynamics}\label{sec:reverse}

For most reversible reactions, rate coefficients are only available for the exothermic (forward) direction. The rate constant for the endothermic (reverse) direction, $k_{r}(T)$ is then calculated as the ratio between the forward rate constant $k_{f}(T)$ and the equilibrium constant $K_{eq}(T)$, calculated with thermodynamical data, as explained in Appendix \ref{Appendix:A}. However, rate coefficients have sometimes been measured for both directions. In such cases, the ratio $k_{f}(T)/k_{r}(T)$ departs from $K_{eq}(T)$ as different uncertainties affect the rate coefficients and the thermodynamic data. The computation of $k_{r}(T)$ using $K_{eq}(T)$ ensures the consistency between kinetics and thermodynamics by making the kinetic model evolve strictly toward the thermodynamic equilibrium that we calculate. Nevertheless, this choice may not always be the best. It results in the propagation into $k_{r}(T)$ of both the errors affecting $k_{f}(T)$ and $K_{eq}(T)$. Indeed, thermodynamic parameters used to calculate $K_{eq}(T)$ are not free of error, and are regularly updated just as kinetic data.  In the field of combustion, for small species as \ce{CH4}, \ce{CH3} and OH, it is common to use experimentally measured kinetic rates, rather than thermodynamical reversal, when they are available in the relevant temperature range. There is no obvious rule in this matter, but validation of the mechanism with out-of-equilibrium experiments seems the only practical way to chose between different rates. This is the criterion that we use and our nominal network uses thermodynamical reversal for most of the reactions but not for three important ones. These reactions affect the unimolecular initiations (or thermal dissociation reactions) of methane into methyl and hydrogen radicals:

\begin{align}
\ce{CH4 ->[M] CH3 + H}\label{reac:CH4}
\end{align}

\noindent of ethane into two methyl radicals:
\begin{align}
\ce{C2H6 ->[M] CH3 + CH3}\label{reac:C2H6}
\end{align}

\noindent and of hydrogen peroxide into two hydroxyl radicals:

\begin{align}
\ce{H2O2 ->[M] OH + OH}\label{reac:H2O2}
\end{align}

At high temperature, chemical kinetics is very sensitive to these three reactions, which have been widely studied experimentally \citep{baulch1994evaluated, golden2008yet, troe2011thermal}. Therefore, we use the kinetic data measured experimentally for the forward and the reverse directions instead of calculating the reverse rate constants using thermodynamic parameters. 

\subsubsection{Excitation of oxygen and nitrogen atoms}

Photodissociations produce excited states of oxygen (O($^1$D)) and nitrogen (N($^2$D)) that are not treated in the original combustion mechanisms. Therefore, we added to the C/H/O/N mechanism, 19 reversible reactions which describe the kinetics of O($^1$D) and N($^2$D), including radiative and collisional desexcitation. These reactions rates are taken (or have been estimated) from \citet{okabe1978photochemistry, herron1999evaluated, umemoto1998, balucani2000cyanomethylene, sato1999measurements, balucani2000observation} and \citet{sander2011}.

\subsection{Test on the chemical scheme with a 0D model}\label{sec:0D}

\begin{figure*}[!htbp]
\centering
{\includegraphics[width=0.95\columnwidth]{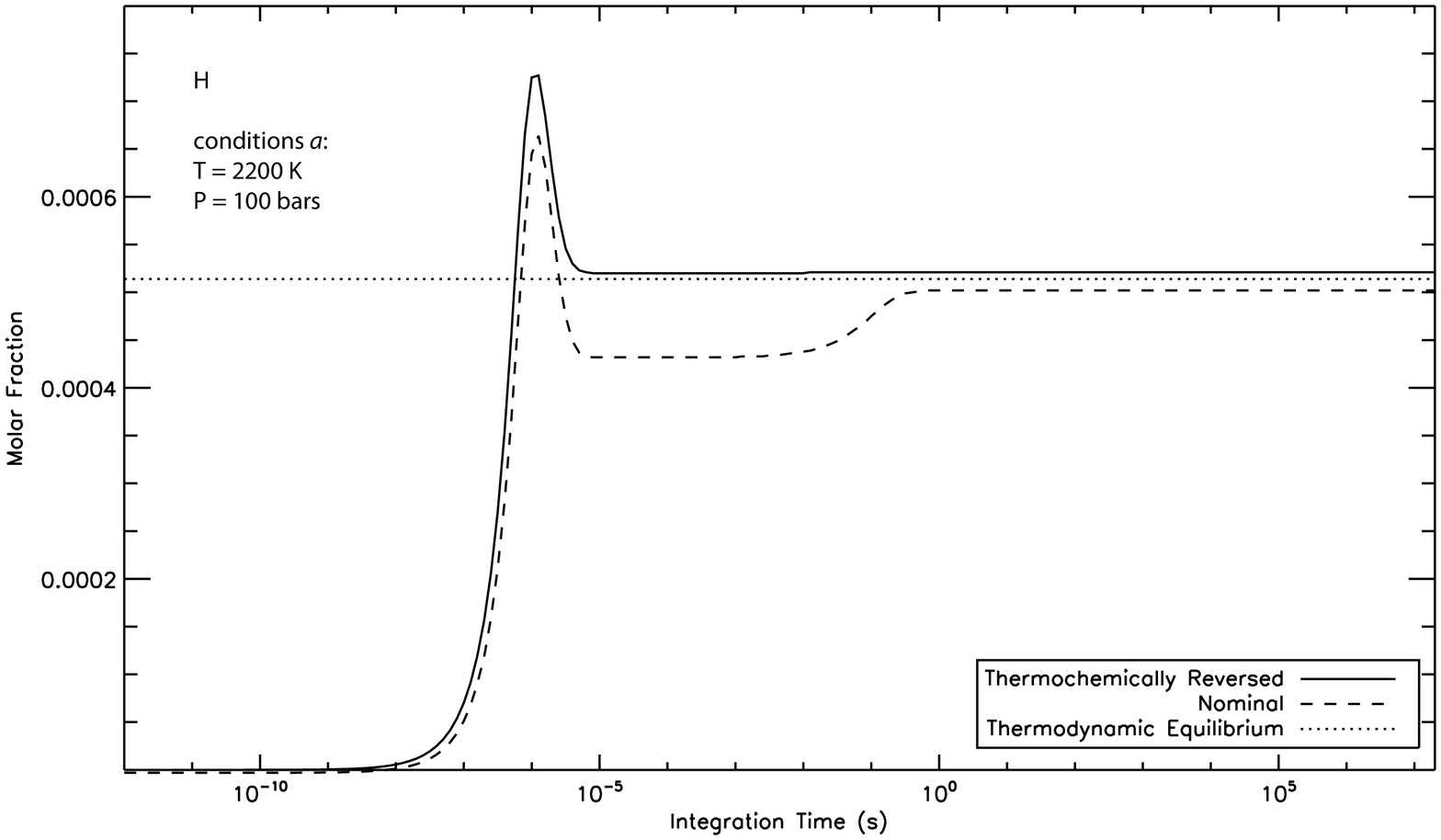}}
{\includegraphics[width=0.95\columnwidth]{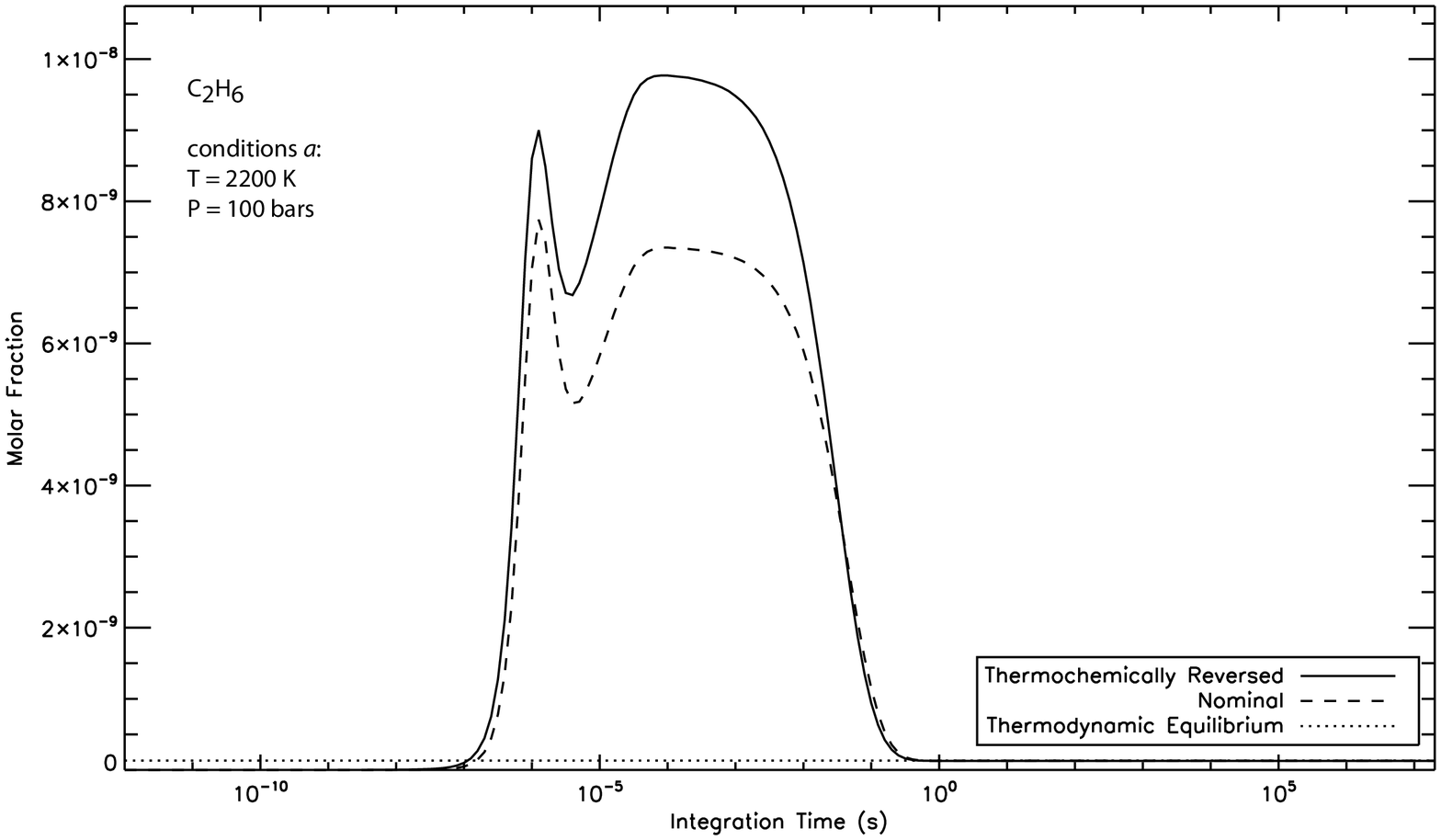}}
{\includegraphics[width=0.95\columnwidth]{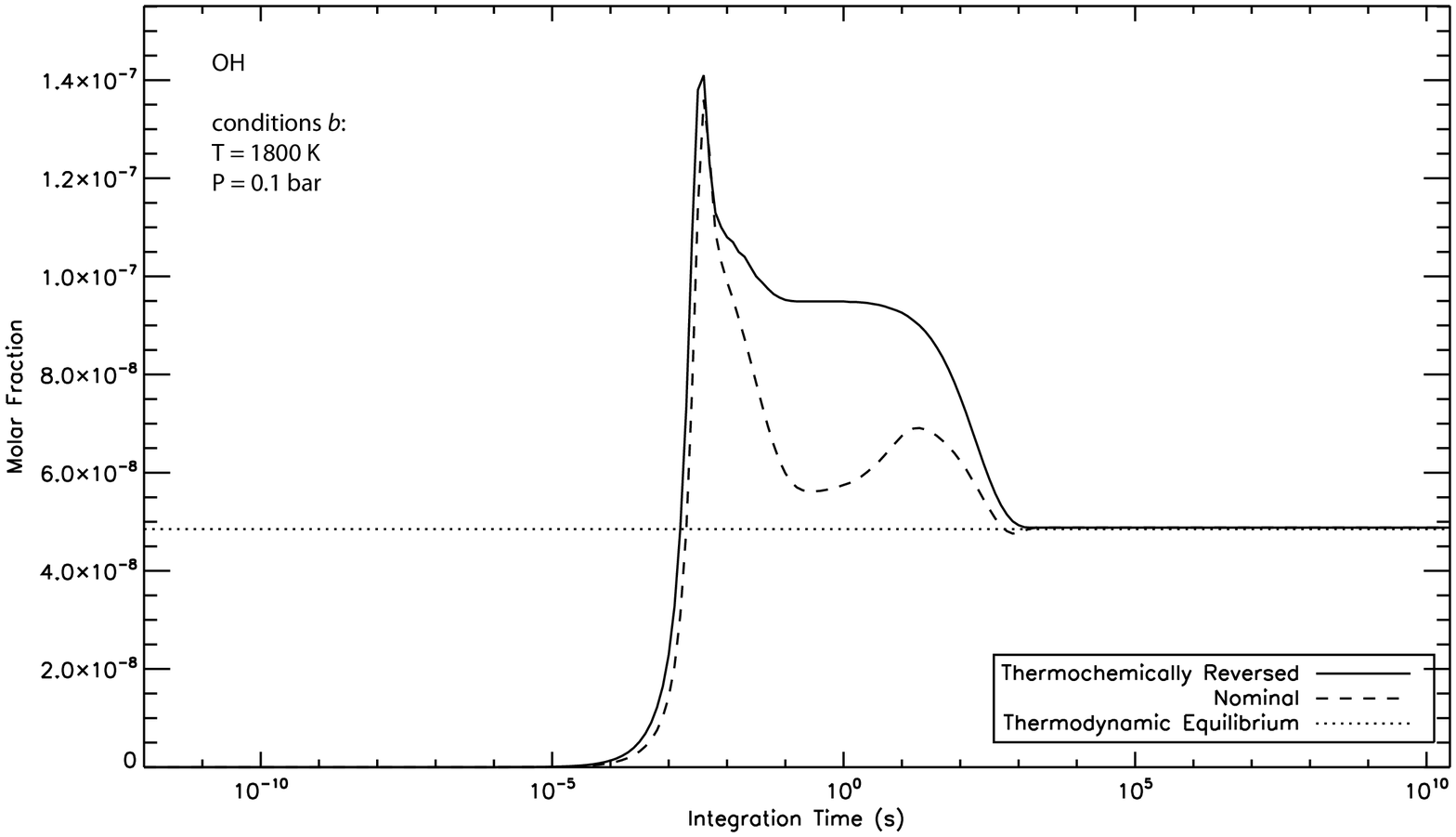}} 
{\includegraphics[width=0.95\columnwidth]{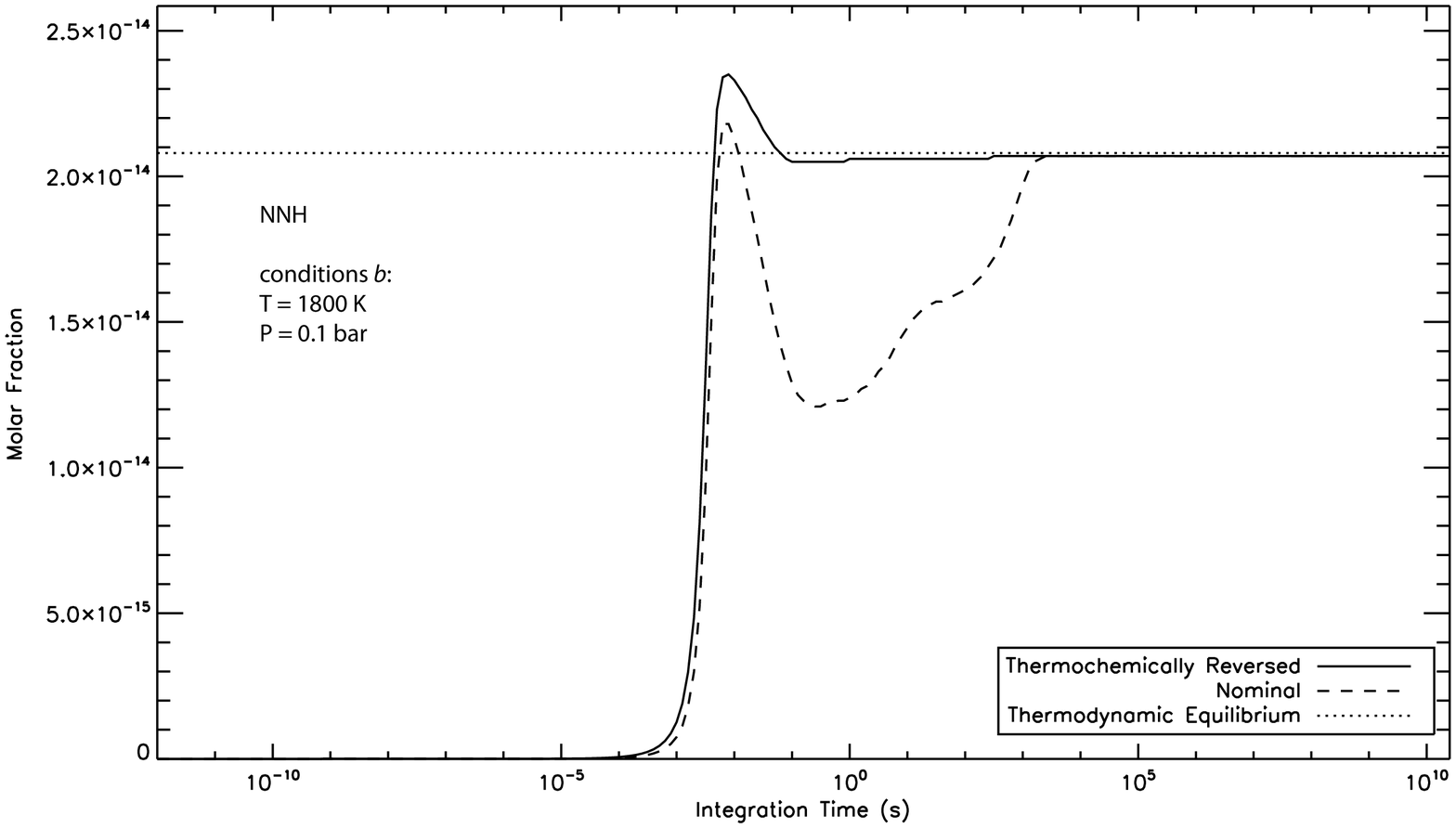}}
\caption{Comparison between the thermodynamic equilibrium (dotted line) and the evolution of some molecular abundances in the 0D model, with two different schemes: the thermochemically reversed model (full line) and the nominal model (dashed line) as a function of integration time at different temperature-pressure points. Initial condition is a mixture of \ce{H2}, \ce{CH4}, \ce{O2}, \ce{N2} and \ce{He} with solar elemental abundances.}
\label{fig:0D}
\end{figure*}

In addition to our 1D model, we have also developed a simple 0D model that computes the chemical evolution of a mixture at constant temperature and pressure. It does not include mixing with another mixture nor photolyses. 
We have used this 0D model to compare the composition found at steady state with the abundances at thermodynamic equilibrium (calculated with the code TECA, described in Appendix~\ref{Appendix:TECA}) for several couples of pressure-temperature. Fig.~\ref{fig:0D} illustrates this comparison with four species. First, we used a version of our nominal scheme in which \textit{all} the reactions are reversed in agreement with their equilibrium constant (solid lines). The computed abundances converge exactly towards the equilibrium values (dotted lines) with negligible numerical differences. Then, we used our nominal model in which some reactions are not reversed according to their equilibrium constant but using rate coefficients measured experimentally . In this case, the abundances reached at steady state (dashed lines) departs from the predicted equilibrium. This departure remains very small: below 1\% for most species, and always below 5\%. However, one can see that the kinetic evolution can be significantly different, both in terms of abundances and timescales.

\subsection{The 1D model}

To model the chemical composition of the atmosphere of the hot Jupiters HD~209458b and HD~189733b, we use our 1D time-dependent model described in \citet{dobrijevic2010key}.
As an input of the model, we give a pressure-temperature profile for the atmosphere of the planet studied. This profile is then divided in discrete uniform layers with a thickness  $\Delta z = \frac{H(z)}{8}$, where $H(z)$ is the pressure scale height. The grid contains $\sim$ 300 layers.
Then the 1D kinetic model resolves the continuity equation (Eq.~\ref{eq:continuite}) as a function of time, for each species and atmospheric layer, until a steady state is reached.

\begin{equation}\label{eq:continuite}
\frac{\partial n_i}{\partial t} = P_i - n_iL_i - div({\Phi_i}\overrightarrow{e_z})
\end{equation}
where $n_i$ the number density of the species $i$ ($\mathrm{cm^{-3}}$), $P_i$ its production rate ($\mathrm{cm^{-3}.s^{-1}}$), $L_i$ its loss rate ($\mathrm{s^{-1}}$),  and $\Phi_i$ its vertical flux ($\mathrm{cm^{-2}.s^{-1}}$) that follows the diffusion equation:

\begin{equation}
\Phi_i = -n_iD_i \left[ \frac{1}{n_i}\frac{\partial n_i}{\partial z}+\frac{1}{H_i}+\frac{1}{T}\frac{dT}{dz}\right]-n_iK\left[\frac{1}{y_i}\frac{\partial y_i}{\partial z}\right]
\end{equation}
where $K$ is the eddy diffusion coefficient ($\mathrm{cm^2.s^{-1}}$), $D_i$ is the molecular diffusion coefficient ($\mathrm{cm^2.s^{-1}}$) and $H_i$ the scale height of the species $i$.\\
At both upper and lower boundaries, we impose a null flux for each species.
\\

\subsection{Photochemistry}

\begin{figure}[!htbp]
\includegraphics[width=0.95\columnwidth]{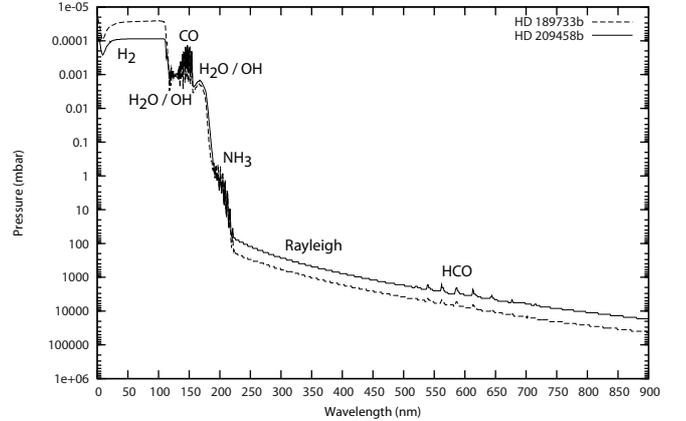}
\caption{Penetration of UV flux in the atmosphere of HD~209458b and HD~189733b at the steady state in function of wavelength. Plots represent the level where the optical depth  $\tau = 1$. The name of the compounds responsible for the main absorption at different wavelengths is indicated.}
\label{penetration}
\end{figure}

We add to the thermochemical scheme a set of 34 photodissociations, which are presented in Appendix~\ref{Appendix:photodisso}.
As we can see in Fig.~\ref{penetration}, for HD~209458b and HD~189733b, UV flux penetrates down to a pressure of about 1 bar, where the temperature is higher than 1500~K.
At these temperature and pressure, endothermic reactions do matter, which implies that photochemistry and thermochemistry are coupled in such highly irradiated atmospheres. We used absorption cross section at the highest available temperature (i.e. 370~K at maximum, which is low compared to the temperatures in the atmosphere of hot Jupiters (see Fig.~\ref{PTprofile})). 

To calculate the photodissociation rates in all the layers of the atmosphere, we compute the stellar UV flux as a function of pressure and wavelength, taking into account molecular absorption by 22 species (Appendix~\ref{Appendix:photodisso}) and Rayleigh scattering. Actinic fluxes are calculated with a resolution of 1~nm (which is also the resolution we adopted for the absorption cross-sections), assuming a plane parallel geometry and an incidence angle $\theta$ of $48^{\circ}$ (as done in \citet{moses2011disequilibrium}, because $<\cos \theta> = 2/3~(\theta \simeq 48^{\circ}$) is the projected-areaweighted average of the cosine of the stellar zenith angle over the planetary disk at secondary-eclipse conditions). Multiple Rayleigh scattering is coupled with absorption through a simple two-stream iterative algorithm \citep{Isaksen1977}.

\subsection{Application to hot Jupiters: HD~209458b and HD~189733b} \label{Sec:application}

HD~209458b and HD~189733b are transiting planets around nearby bright stars. Their atmospheres have been studied by their transmission spectrum obtained during the primary transit and their day-side emission spectrum measured at the secondary eclipse. These observations can be used to constrain the thermal profile \citep{swain2009water, madhusudhan2009temperature} and to detect the spectral signature of atmospheric compounds \citep{charbonneau2002detection, tinetti2007water, swain2008presence, grillmair2008strong, langland2009hires, swain2009water, swain2009molecular, beaulieu2010water}.

In this preliminary study, we do not compare the results of our model with these observations for two reasons. First, because there is not yet a consensus on the actual constraints that can be drawn from these measurements. Secondly, such a comparison would imply to address the effects of circulation on the composition and to explore all the range of possible elemental abundances for these objects. Several recent works claim that observations of some hot Jupiters imply enhanced elemental C/O ratios \citep{madhusudhan2011high, madhusudhan2011carbon}. With C/O ratios close or above unity, species with more than two carbon atoms will be important and our C$_0$-C$_2$ network does not allow us to study them. For these reasons, we are implementing a C$_0$-C$_6$ mechanism and a coupling with atmospheric circulation, which will be described in further studies. At this stage, our main goal is to compare the results of our model with already published works, in particular \citet{moses2011disequilibrium}, hereafter M11. We could also have compared our results with those of \citet{zahnle2009atmospheric}, who explored a broader range of conditions, included sulfur-bearing species and various elemental compositions. We decided to restrict our comparison with M11 because their model, like ours, only includes species made of C, H, O and N (and He), and also because M11 already did a comparison between their results and those of \citet{zahnle2009atmospheric} showing only little discrepancies when the same conditions are considered. We used the same conditions ($P$-$T$ profiles, eddy diffusion, elemental abundances) as in M11, so that differences should come only from kinetics (and photochemistry in the upper atmosphere), which represents the novelty of our approach.

\subsubsection{Physical properties and composition}

The physical properties  of HD~209458b have been refined by \citet{rowe2008very} and are presented in Table~\ref{properties}, with some properties of the host star. Properties of HD~189733b and HD~189733 come from \citet{southworth2008homogeneous, southworth2010homogeneous}.

\begin{table}[!htbp]
 \caption{Properties of the systems HD~209458 and HD~189733 }\label{properties}
\centering
\begin{tabular}{lcc}
\hline
\hline
 &HD~209458 & HD~189733\\
\hline
Distance Sun-Star (pc)  & 47 & 19.3 \\
Distance Planet-Star (AU)  & 0.047 & 0.03142\\
$M_p (M_{Jup})$ & $0.69 \pm 0.01$ & $1.150 \pm 0.028$ \\
$R_p (R_{Jup})$ & $1.339\pm0.002$& $1.151\pm0.036$\\
$\rho_p (\rho_{Jupiter})$ & $0.26 \pm 0.04$& $0.755 \pm 0.066$\\
$g_0 \mathrm{(m.s^{-2})}$ & $9.54\pm 0.69$ &$21.5 \pm 1.2$\\%$(9.30 \pm 0.08)$ avec southworth 2010
$P_{orbit}$ (days) & 3.5247489(2) & 2.21857578(80) \\
$M_\star (M_\odot)$ & $1.083\pm 0.005$ & $0.840\pm 0.030$\\
$R_\star (R_\odot)$ & $1.118\pm0.002$ & $0.752\pm0.023$\\
Spectral type & G0 V & K1 V - K2 V \\
\hline
 \end{tabular}
 \tablenotea{The 1$\sigma$ uncertainty in $P_{\rm orbit}$ is given in parentheses in units of the last digits}
 \end{table}
 
\begin{figure*}[!htbp]
\includegraphics[width=0.95\columnwidth]{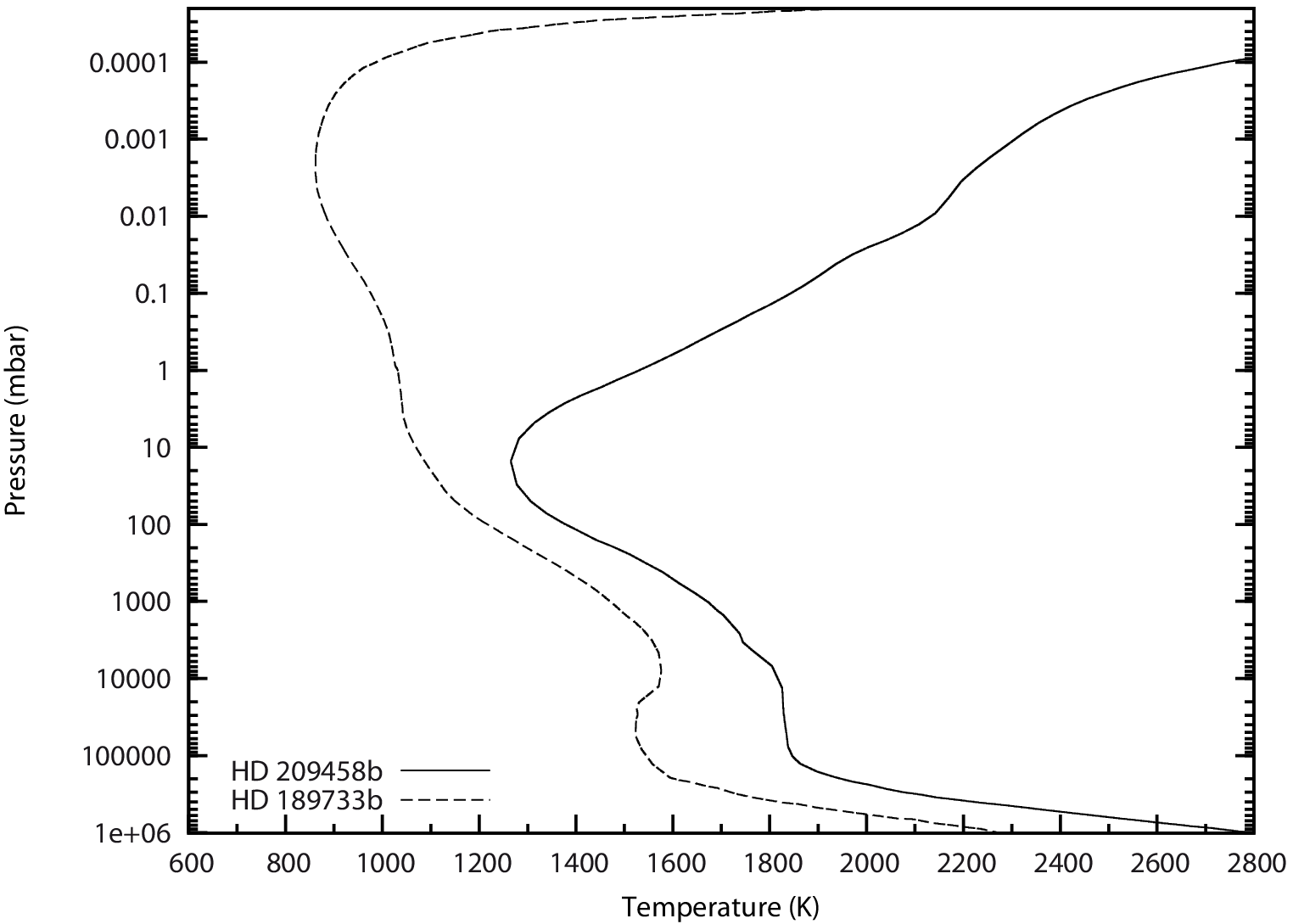}
\includegraphics[width=0.95\columnwidth]{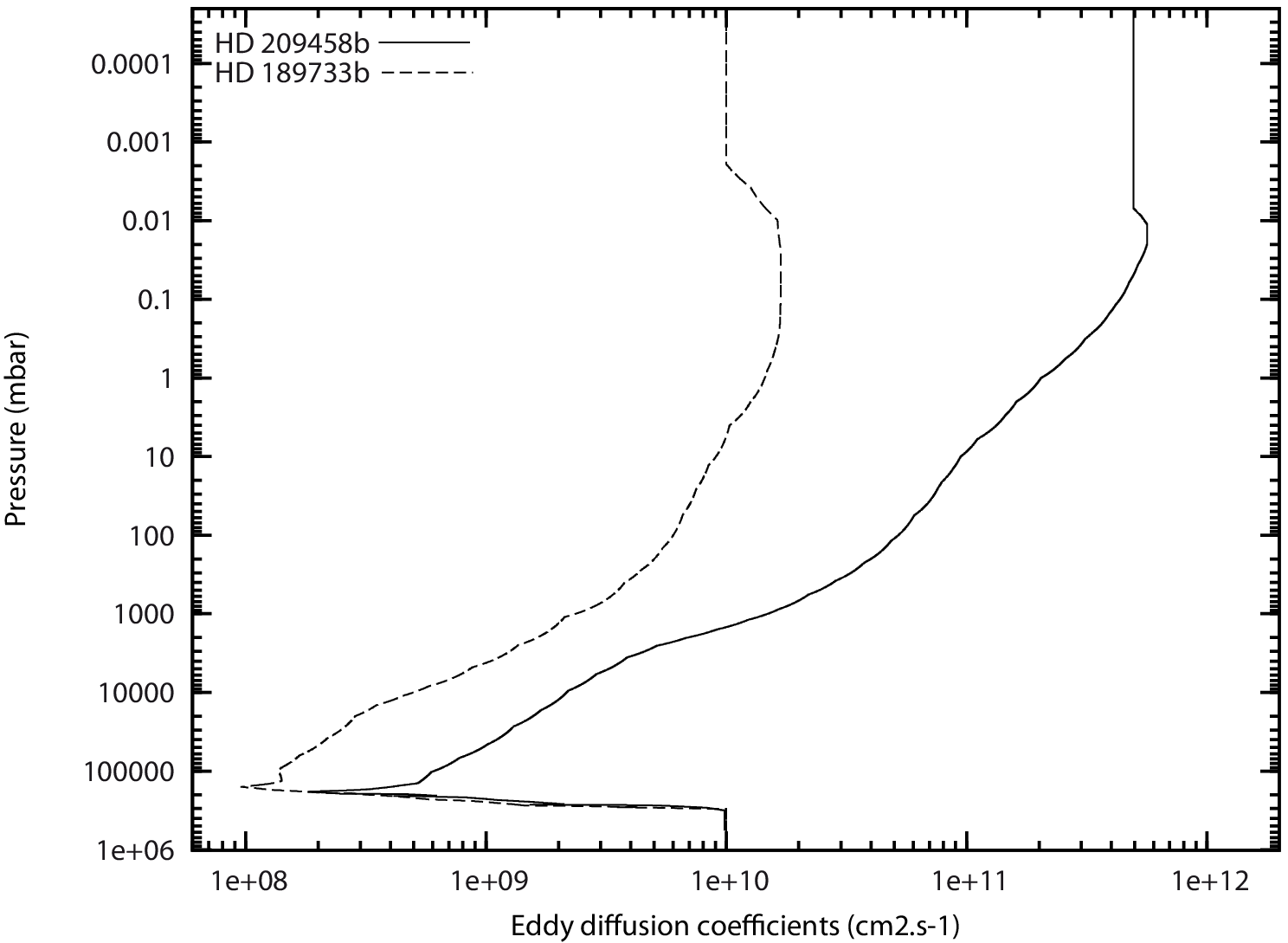}
\caption{Pressure-temperature profiles (left) and eddy diffusion profiles (right) of HD~189733b and HD~209458b (from \citealt{moses2011disequilibrium}).}
\label{PTprofile}
\end{figure*}

In order to compare the outcomes of the two models (Sect.~\ref{sec:comp_moses}), we use the temperature and eddy diffusion profiles published in M11 (Figs.~\ref{PTprofile}). Also following M11, we assume protosolar elemental abundances \citep{lodders2009solar} for both planets, with 20\% of depletion for oxygen (sequestered along with silicates and metals). We start our time-dependent modeling with the thermodynamic equilibrium abundances calculated with TECA (an equilibrium model described in Appendix~\ref{Appendix:TECA}) at each level of the atmosphere.

\subsubsection{UV spectral irradiance}

As HD~209458 is a G0 star (Table~\ref{properties}), we use the UV spectral irradiance of the Sun for this star.
For the star HD~189733, a K1~-~K2 star (Table~\ref{properties}), the UV  spectrum has been provided to us by Ignasi Ribas (private communication). It is based on FUSE and HST observations of the star $\epsilon$ Eridani, a proxy of HD~189733 (similar type, age and metallicity), in the 90-330 nm range. Between 0.5 and 90~nm, it is based on data from the X-exoplanets Archive at the CAB \citep{sanz-forcada2011}. Above 330 nm, we use a synthetic spectrum calculated with the stellar atmosphere code Phoenix \citep{hauschildt1999nextgen}. This model for the UV spectrum of HD~189733 slightly differs from the one chosen in M11. We also tested our model with the spectrum used by M11 and found negligible differences at the pressure levels we model.
\\

\section{Results}\label{sec:results}

\subsection{Nominal model}

First of all, we checked that our kinetic model reproduces the thermodynamic equilibrium, in the absence of vertical mixing and photodissociation. We obtained differences lower than a few percent, as found with the 0D model (see Sect.~\ref{sec:0D}). 
For both planets, the homopause is always found above the $1\times10^{-5}$ mbar level, which is beyond the range of pressure that we model. As a consequence, and although it is included, molecular diffusion does not affect our results. 

\begin{figure*}[!htbp]
\includegraphics[width=0.95\columnwidth]{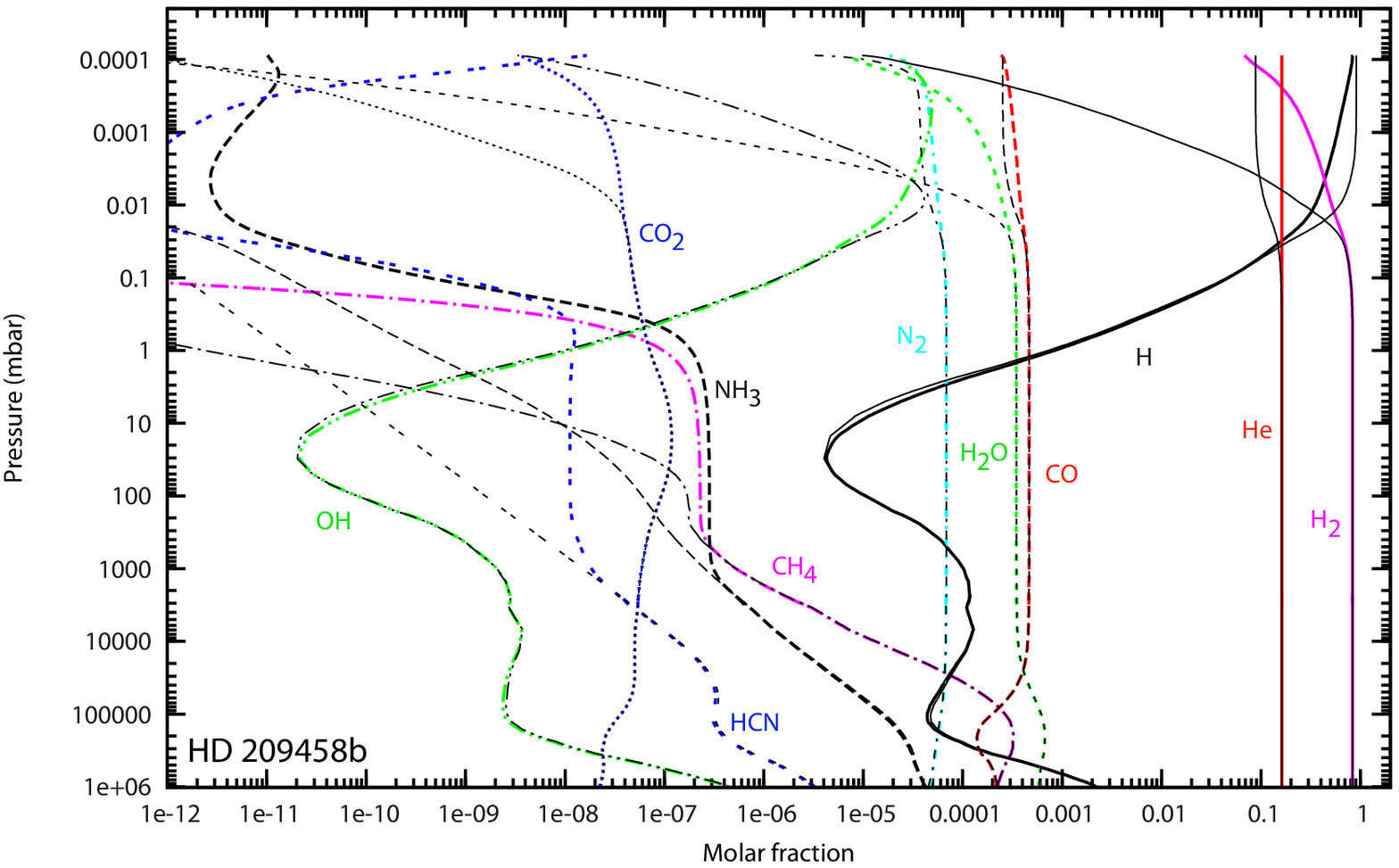}
\includegraphics[width=0.95\columnwidth]{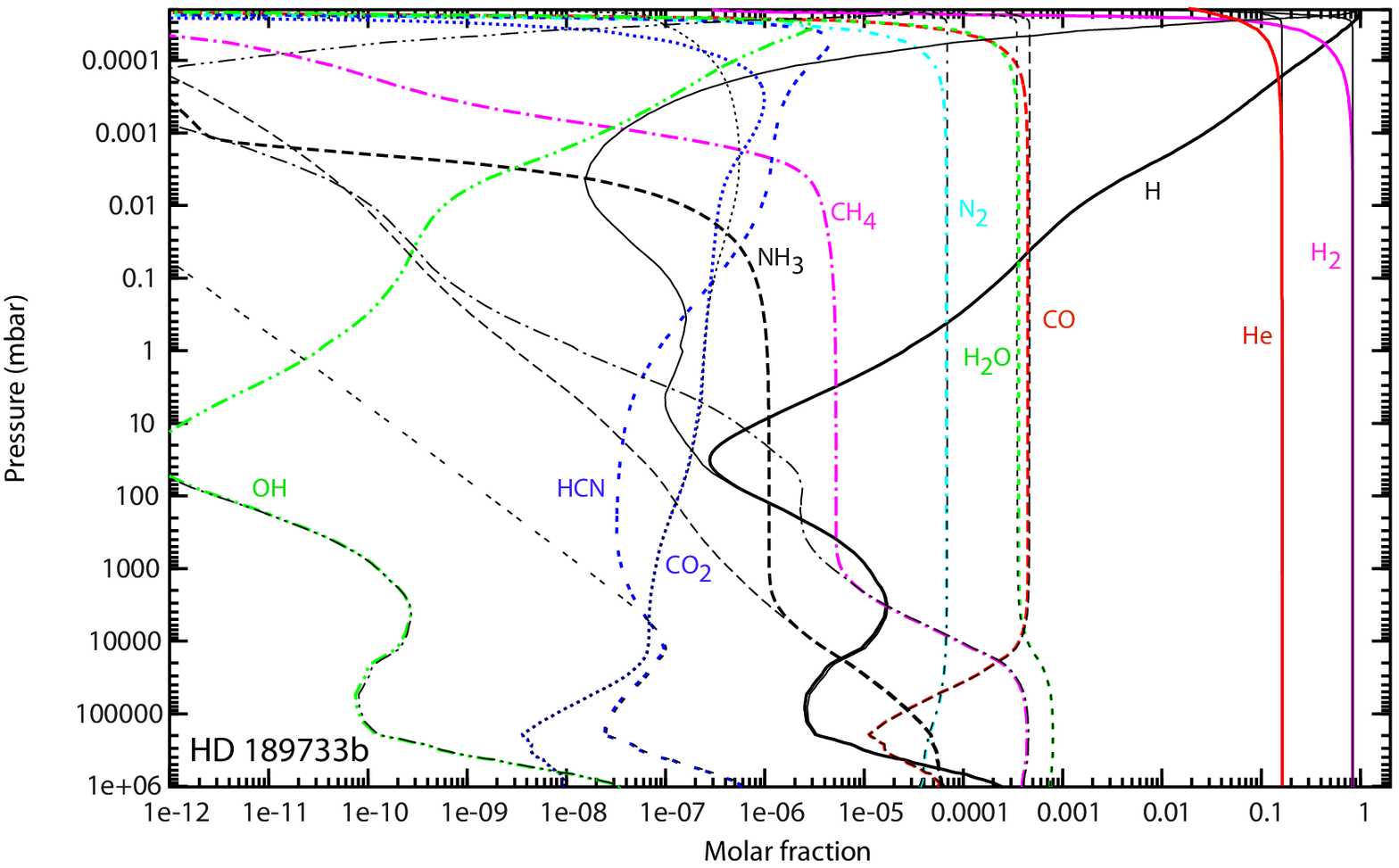}
\caption{Steady-state composition of HD~209458b (left) and HD~189733b (right) calculated with our nominal model (color lines), compared to the thermodynamic equilibrium (thin black lines).}
\label{results}
\end{figure*}

\begin{figure*}[!htbp]
\includegraphics[width=0.95\columnwidth]{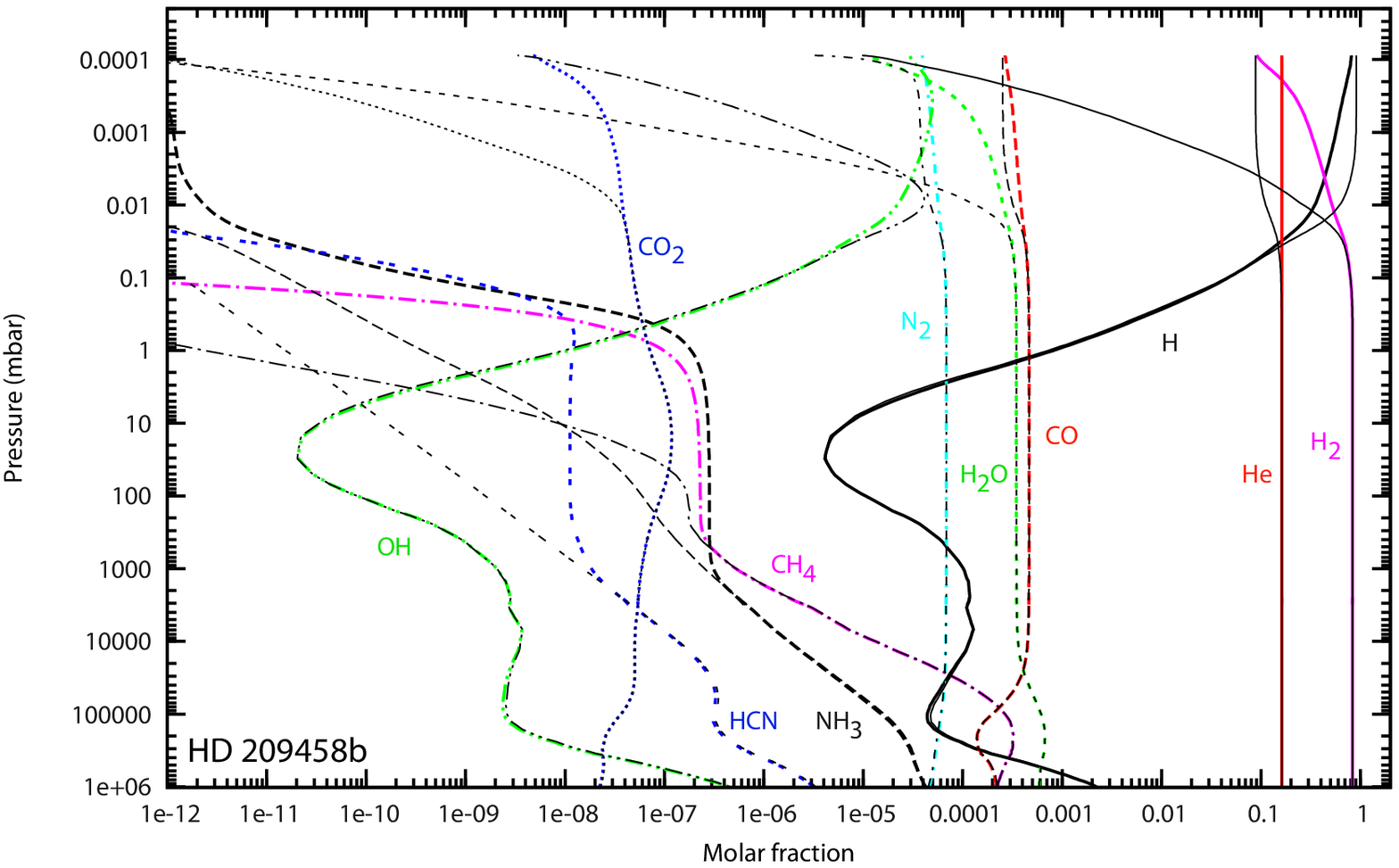}
\includegraphics[width=0.95\columnwidth]{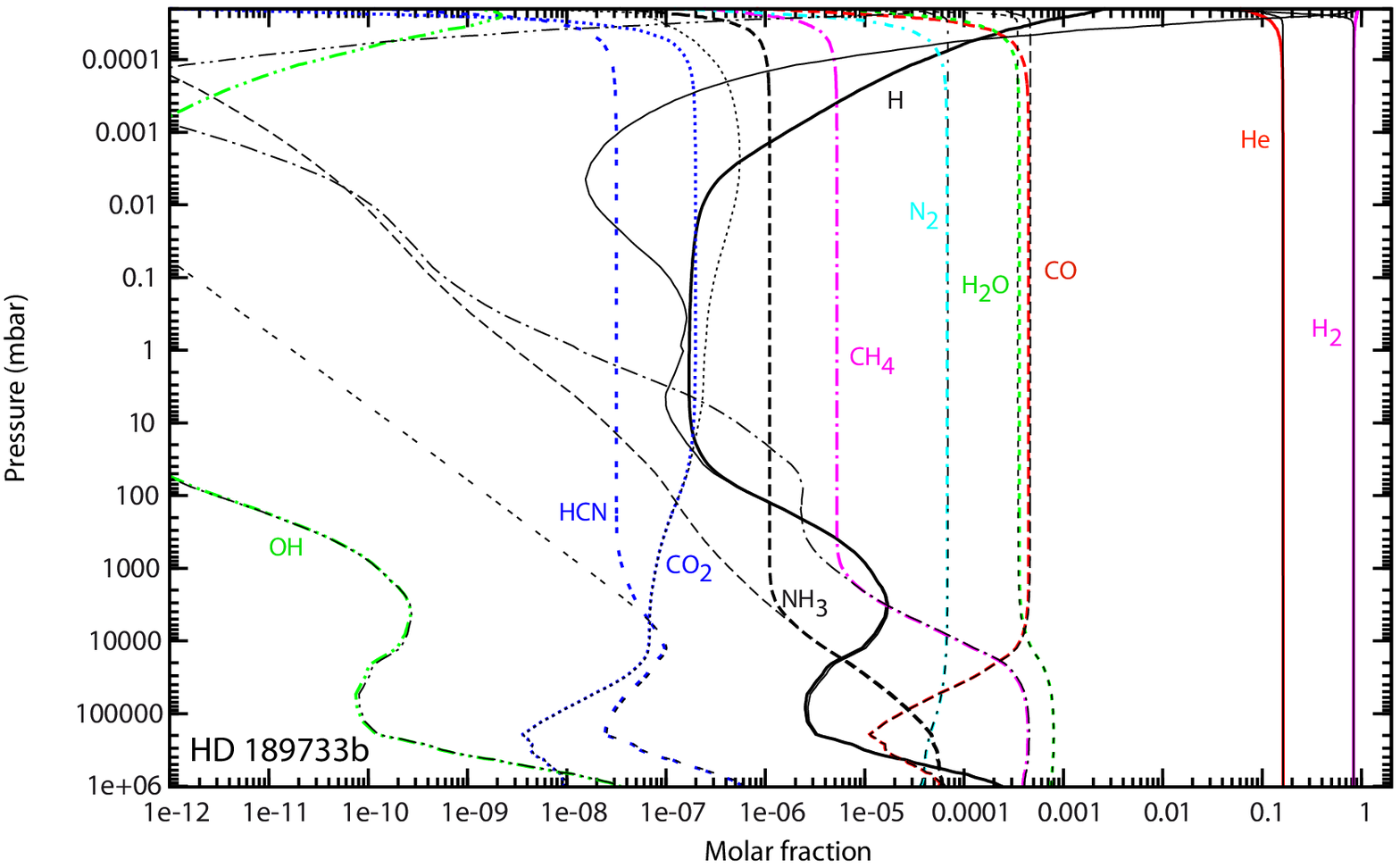}
\caption{Steady-state composition of HD~209458b (left) and HD~189733b (right) calculated with our nominal model without photodissociation (color lines), compared to the thermodynamic equilibrium (thin black lines).}
\label{sanshv}
\end{figure*}

Figure~\ref{results} shows the steady-state composition of the atmosphere of HD~209458b and HD~189733b, with vertical transport and photodissociations, while in Fig.~\ref{sanshv}, photodissociations have been removed.
Comparing Figs.~\ref{results} and \ref{sanshv} shows us the influence of photolyses. Although HD~209458b receives a higher UV flux than HD~189733b, we can see that UV photons have little effect on the composition of HD~209458b, while they have a significant influence on the chemistry of HD~189733b. This is because the temperature is higher in HD~209458b so that the chemical timescales are significantly shorter than the lifetime of species against photolyses. So in HD~209458b, kinetics dominate over photodissociations, even at high altitude. 
In HD~189733b, however, photodissociations affect the composition down to about the 10 mbar level. This is particularly noticeable for H and OH abundances.
The production of H is dominated by the photolysis of \ce{H2} for pressures lower than 1 $\mu$bar. Below this level, and for pressures higher than 0.1 mbar, H is produced by the photodissociation of \ce{H2O}, with a minor contribution of the photodissociations of \ce{NH3} and \ce{HCN}.
%Vertical mixing produce a smooth abundance profile through these photochemical production zone and the thermochemical zone. 
The abundance of OH follows the profile of H, and increases for pressures lower than 10 mbar. There is a photochemical enhancement of HCN above the 10 mbar pressure level, as discussed in M11.
\ce{CH4} is destroyed by photolyses for pressures lower than 0.01 mbar. \ce{NH3} is photodissociated down to levels as deep as 1 bar, but vertical transport compensates this destruction for pressures higher than 0.1 mbar. Above that level, the amount of \ce{NH3} decreases with altitude due to photolyses. Photochemistry has a negligible effect on \ce{CO2}, as noted by \citet{zahnle2009atmospheric}.

For HD~209458b, we can see in Fig.~\ref{sanshv} that mixing quenches \ce{NH3} and \ce{HCN} at 1 bar and \ce{CH4} at 400 mbar. These species are transported up to $\sim$1 mbar pressure level, but as the temperature increases with altitude at this level, they tend to come back to their thermochemical equilibrium values, so their abundances decrease again. For the other molecules, like \ce{H2}, H and \ce{CO2}, at the thermodynamic equilibrium, there is a steep variation of composition (smoothed by vertical mixing) corresponding to the strong temperature gradient of the upper atmosphere.\\
Vertical quenching has an effect on a larger part of the atmosphere of HD~189733b. \ce{NH3} and \ce{HCN} are quenched at 5 bar, \ce{CH4} at 1 bar, H at 40 mbar and \ce{CO2} at 20 mbar. Quenching contaminates the composition up to very low-pressure levels ($10^{-4}$~mbar). 

\subsection{Comparison with Moses et al. (2011)}\label{sec:comp_moses}

\subsubsection{Equilibrium}
Overall, the composition we calculate at thermodynamic equilibrium (which is our initial condition) is very close to what is obtained by M11 except for HCN for which there is a difference that can reach $\sim$ 30\% at 100 bars and 1545 K. To check our calculations we also did a comparison with the code STANJAN\footnote{http://navier.engr.colostate.edu/tools/equil.html} and we found negligible discrepancies for the species we compared, including HCN at this pressure and temperature. The difference with M11 probably comes from the coefficients used for the NASA polynomials (see Appendix~\ref{Appendix:A}). Although the difference remains small for equilibrium calculations, we should keep in mind that it may significantly affect the kinetics of HCN and related species through the calculation of the rate for backward reactions and vertical quenching.

\subsubsection{Steady-state}

\begin{figure*}[!htbp]
\includegraphics[width=0.95\columnwidth]{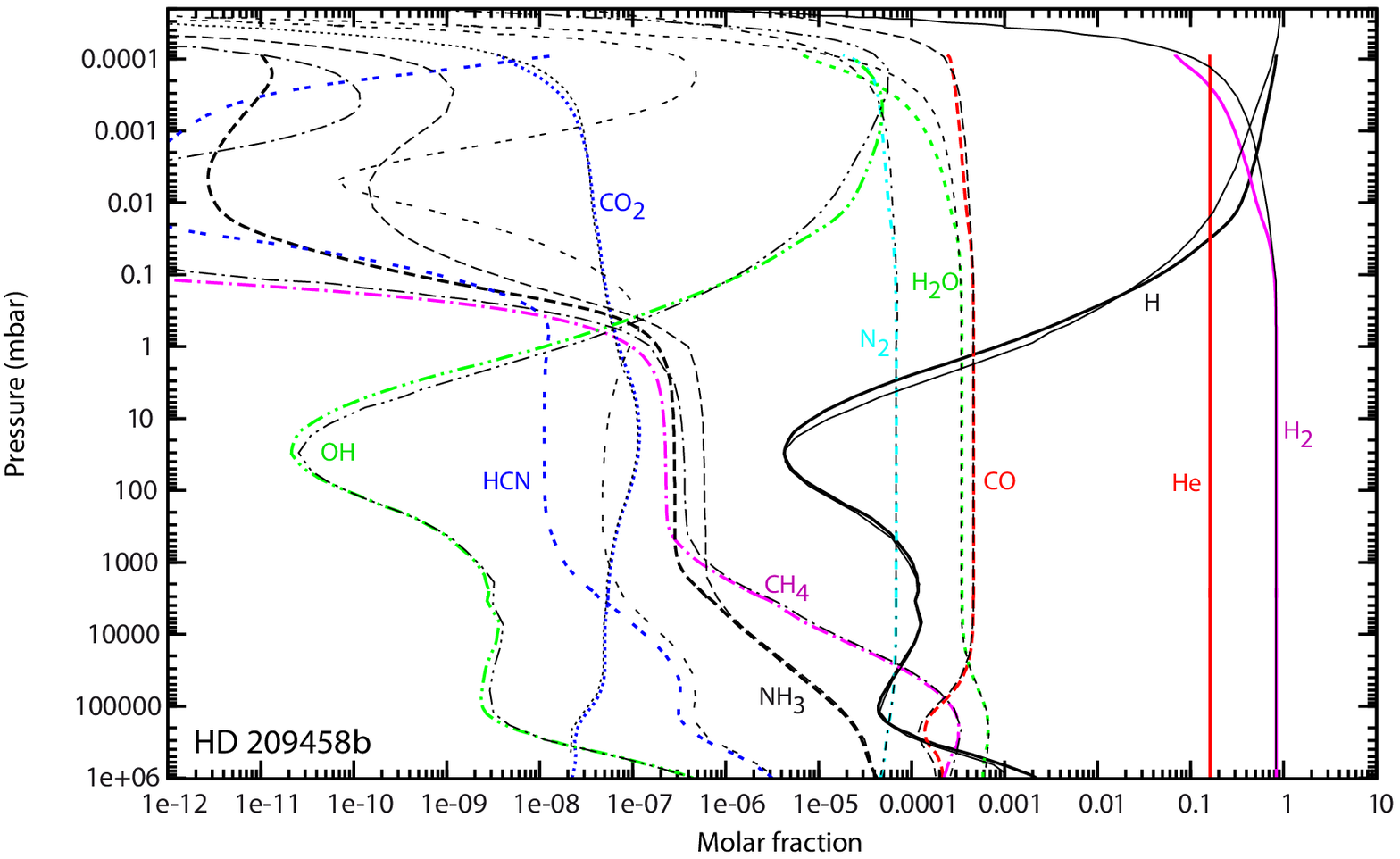}
\includegraphics[width=0.95\columnwidth]{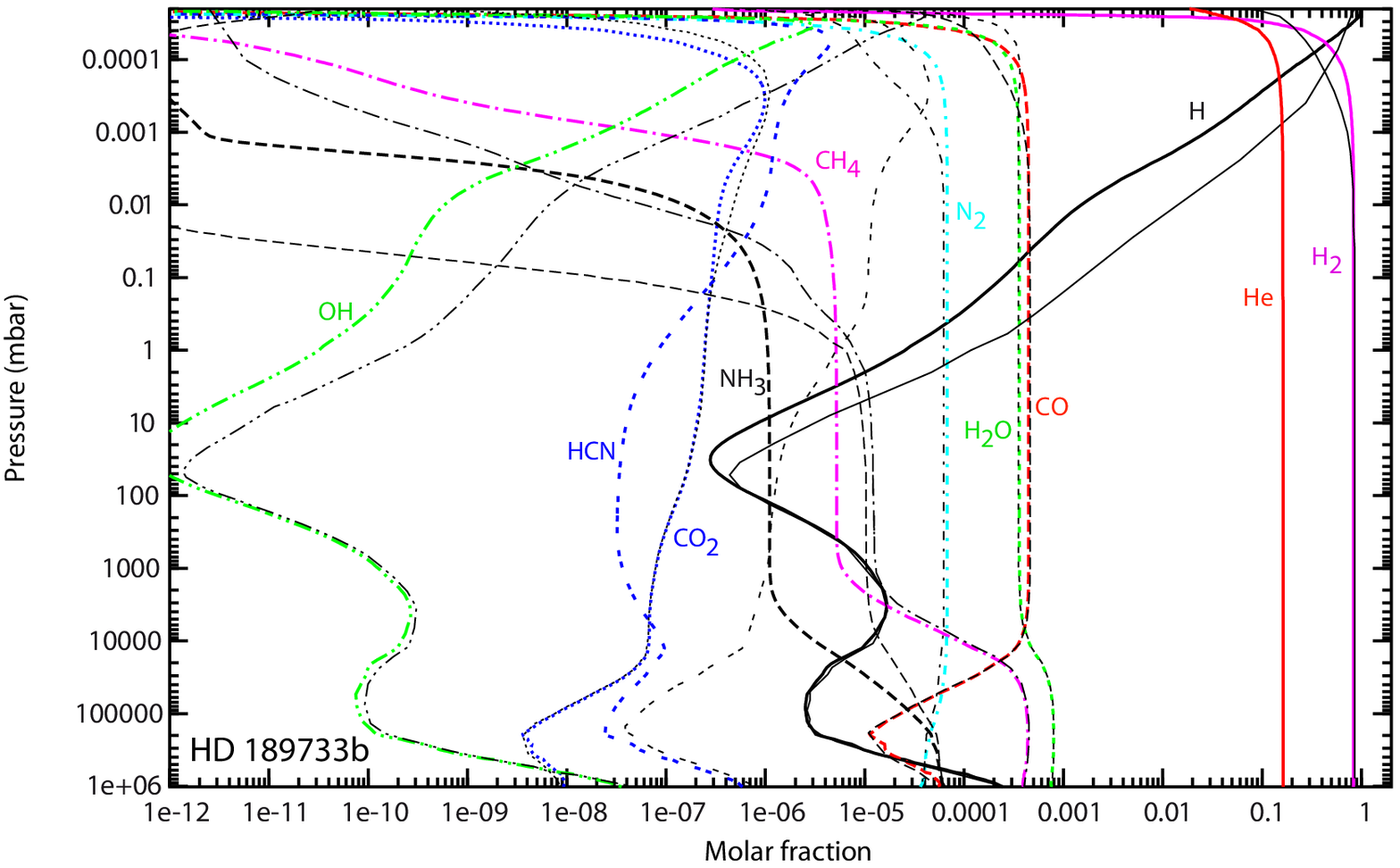}
\caption{Comparison between the abundance profiles found by our nominal model (color lines) and by \citet{moses2011disequilibrium} (thin black lines), for the two planets (HD~209458b (left) and HD~189733b (right))}
\label{comp_moses}
\end{figure*}

In Fig.~\ref{comp_moses}, we compare our results at steady state with those of M11. Differences between M11 and our nominal model are also shown species by species in Figs.~\ref{fig:models1} and \ref{fig:models2}. Discrepancies between the two models can be due to the different chemical schemes and, at levels where the results are sensitive to photochemistry, to possible differences in the UV fluxes, cross sections, and photodissociation quantum yields. An influence of the numerical implementation (like the discretization of the atmosphere, the solver for the continuity equations, or the treatment of the UV transfer) is also possible. 

In the lower atmosphere of HD~189733b and for most of the atmosphere of HD~209458b, photolyses have a negligible influence and departures should be caused by the kinetics. For these regions we find very similar results for species that remain at their equilibrium abundance (H, OH, \ce{CO}, \ce{CO2}, \ce{H2O} for instance), which only confirms, as stated before, that our thermodynamic equilibrium codes are in good agreement.  For species quenched by mixing, however, significant deviations appear, in particular for \ce{NH3}, HCN and \ce{CH4}. Their quenching occurs at different pressure levels and, thus, for different abundances that will then contaminate a large fraction of the atmosphere above. The discrepancies are much more significant in the case of HD~189733b, due to a higher sensitivity to kinetics. Although the kinetic network is certainly the main reason for these departures, it is also true that quenching can be quite sensitive to the resolution of the pressure (or altitude) grid, in particular when there is a steep gradient of temperature which is the case in the convective zone ($P>$100~bar). For this reason, we impose the thickness of individual layers to be smaller than $1/8^{th}$ of the local scale height, which results in $\sim$300 layers for the pressure range that we model. Although we do not know what resolution is used in M11, it seems more likely that the deviation comes from differences in the kinetic network itself. As explained in the Introduction, we use a chemical scheme validated for the species represented and for most of the range of temperature and pressure of the modeled atmospheres. M11, on the other hand, use a chemical scheme derived from Jupiter and Saturn models \citep{gladstone1996hydrocarbon, moses1996, moses1995post, moses1995nitrogen, moses2000photochemistry, moses2000photochemistry2} completed by high-temperature kinetics from combustion-chemistry literature \citep{baulch1992evaluated, baulch1994evaluated, baulch2005evaluated, atkinson1997evaluated, atkinson2006evaluated, smithgri, tsang1987chemical, tsang1991chemical, dean2000gas}, which has not, to our knowledge, been validated against experiments.
We also note departures in the upper atmosphere, where photolyses are important. In particular, H and OH exhibit similar profiles than those of M11 in HD~189733b but shifted by about one order of magnitude in abundance for pressures lower than 50~mbar. \ce{CH4} is also affected. We checked that these differences are not due to the use of different stellar fluxes by switching between the flux we use and the one used in M11 (for HD~209458b, we both use the solar UV flux). At these altitudes, we note a significant sensitivity of the mixing ratio of these species to the Rayleigh scattering, so the treatment of the scattering could explain at least part of this disagreement. Again, and although we do not know the details of the photochemical data and radiative transfer used in M11, we assume that kinetics explain the differences.

\subsection{Other networks for nitrogen species} \label{sec:nitrogen2}

The main differences between M11 and our results is related to the quenching of \ce{NH3} and \ce{HCN}. As mentioned in M11, the chemistry of nitrogen compounds has been less studied than carbon species and chemical networks have been subjected to less validation. However, NO$_{X}$, HCN, CN and \ce{NH3} are important species in applied combustion (gas fuel, for instance, can contain high concentrations of ammonia), and should be well reproduced within the temperature and pressure range of the validation. Quenching is found to occur within 1 to 10 mbars, corresponding to the range of validation in terms of pressure. %(experiments done above ten bars are scarce).
An originality of our network compared to other schemes used in combustion is that it is not optimized to increase the agreement between modeling and experiments. In other words, the rate coefficients of the individual processes have not been altered compared to their original measurement or estimate. The application of the network is therefore not strictly restricted to the validation domain.\\
Other sub-mechanisms are available to model the kinetics of nitrogen-bearing species. They have been constructed based on different approaches (optimization, specific domain of application, reduced number of reactions). In order to test our model against other nitrogen schemes, we replaced our nitrogen reaction base by nitrogen sub-mechanisms taken from other C/H/O/N mechanisms: 

- {\it GRIMECH}, mechanism based on GRI-Mech 3.0 \citep{smithgri} with several reactions involving NO$_X$ compounds added with respect to the mechanisms of \citet{glaude2005kinetic} as recommended and done by \citet{Anderlohr2009505}. It includes 162 reversible reactions involving 26 nitrogen compounds. The GRI-Mech 3.0 is a mechanism designed to model natural gas combustion, including NO formation and reburn chemistry. As already mentioned in Sect.~\ref{sec:nitrogen}, it has been optimized as a global mechanism, i.e., some rate coefficients have been modified (relatively to the literature) in order to fit the results of a pool of experiments with conditions and compositions specific to combustion. The individual processes have not been studied separately in all the pressure and temperature range. Applying this mechanism beyond its domain of optimization/validation is a risky extrapolation. Mixing ratios of oxidants, for instance, are very low in hot Jupiter atmospheres compared with the experiments used to optimize/validate GRI-Mech 3.0.\\
While doing the present study, we noticed that two reactions from the {\it GRIMECH} mechanism had wrong rates and corrected them.
These erroneous rate constants, that can be traced back to the NIST Chemical Kinetics Database \citep{NIST}, were identified by systematically comparing reaction rate constants with collision limit values and energy barriers with the enthalpy budget of the reaction.

\noindent The first reaction is

\begin{align}\label{reac:N2}
\ce{N2 + H -> N(^4S) + N(^4S) + H}
\end{align}

\noindent for which the rate given by NIST is $k_f(T)$=1.26 $\times$ 10$^{-9}$ $T^{-0.20}$e$^{-27,254/T}$ cm$^3$ molecule$^{-1}$ s$^{-1}$ (with $T$ in Kelvin), although this expression corresponds in fact to the reaction \ref{reac:NH}, the thermal dissociation of NH through collisions with atomic nitrogen \citep{caridade2005unimolecular}: 

\begin{align}\label{reac:NH}
\ce{NH + N -> H + N2}
\end{align}

\noindent The above expression overestimates by many orders of magnitude the rate constant of the reaction, whose activation energy must be around 100,000 K, as implied by the bond energy of molecular nitrogen and by measurements of the thermal dissociation of N$_2$ through collisions with various bodies. 

\noindent We finally adopted for the dissociation of \ce{N2} a more general form:

\begin{align}\label{reac:N2M}
\ce{N2 + M -> N(^4S) + N(^4S) + M}
\end{align}

\noindent with the reaction rate constant  $k_f(T)$=1.661 $\times$ 10$^{4}$ $T^{-3.30}$e$^{-11,310/T}$ cm$^3$ molecule$^{-1}$ s$^{-1}$ \citep{thielen1986n}. This rate had a strong influence on our results. Note that using the wrong rate has a large effect on the atmospheric profiles of  \ce{NH3} and  HCN.

\noindent The second reaction is \\

\begin{align}\label{reac:HONO}
\ce{HONO + NO -> NO2 + HNO}
\end{align}

\noindent for which the reaction rate constant given by NIST ($k_f(T)$=7.34 $\times$10$^{-20}$ $T^{2.64}$ e$^{-2034/T}$ cm$^3$ molecule$^{-1}$s$^{-1}$) was in fact the rate of the reverse of reaction (\ref{reac:HONO}), as calculated by \citet{mebel1998}. In our model, we use in fact the reaction :

\begin{align}\label{reac:NO2}
\ce{NO2 + HNO -> HONO + NO}
\end{align}

\noindent with the reaction rate constant $k_f(T)$=1.00 $\times$10$^{-12}$ e$^{-1000/T}$ cm$^3$ molecule$^{-1}$s$^{-1}$ \citep{tsang1991chemical1}.

\noindent When this paper is published, these rates should be corrected in the NIST Chemical Kinetics Database, but one should check that the wrong rates are not used in modeling. 

- {\it GDF-Kin}, a mechanism optimized for natural gas combustion modeling \citep{turbiez1998gdf, de2008experimental} that includes less individual processes: 180 reversibles reactions involving 22 nitrogen species. Several experimental data on natural gas combustion have been acquired in partnership with \textit{Gaz de France} to develop this mechanism. NO$_{X}$ chemistry has been included in GDF-Kin 3.0 \citep{Elbakali2006896}. We use the update version GDF-Kin 5.0 \citep{lamoureux2010experimental}, in which 5 reactions involving NCN have been refined in order to better reproduce the kinetics of this species. It is validated for temperatures between 400 and 2200~K and pressures between 0.04 and 10~bars.

- {\it DEAN}, taken from \citet{dean2000gas}. This book that presents a catalog of reactions is used by \citet{moses2011disequilibrium}, at least for some reactions. The mechanism derived from this work  includes 370 reversible reactions involving 49 nitrogen species and one C/H/O species that is not included in our C$_0$-C$_2$ scheme: HCOH. The purpose of the work of Dean and Bozzelli was to list gas phase reactions involving nitrogen-bearing species that could be important for high temperature combustion modeling and to provide the associated rate coefficients based on an analysis of elementary reaction data, when available, or on estimations from thermochemical kinetics principles otherwise. This mechanism was developed on the basis of analysis of individual reactions rather than by attempting to reproduce any specific set of experiments. It is clearly written in this book that: "Although we show in the chapter that this mechanism provides a reasonable description of some aspects of high-temperature nitrogen chemistry, we have not attempted a comprehensive comparison". Therefore, this kinetic network should be viewed as a database of reaction rate constants rather than a validated mechanism, in the absence of validation.

\begin{figure*}[!htbp]
\centering
\includegraphics[width=\columnwidth]{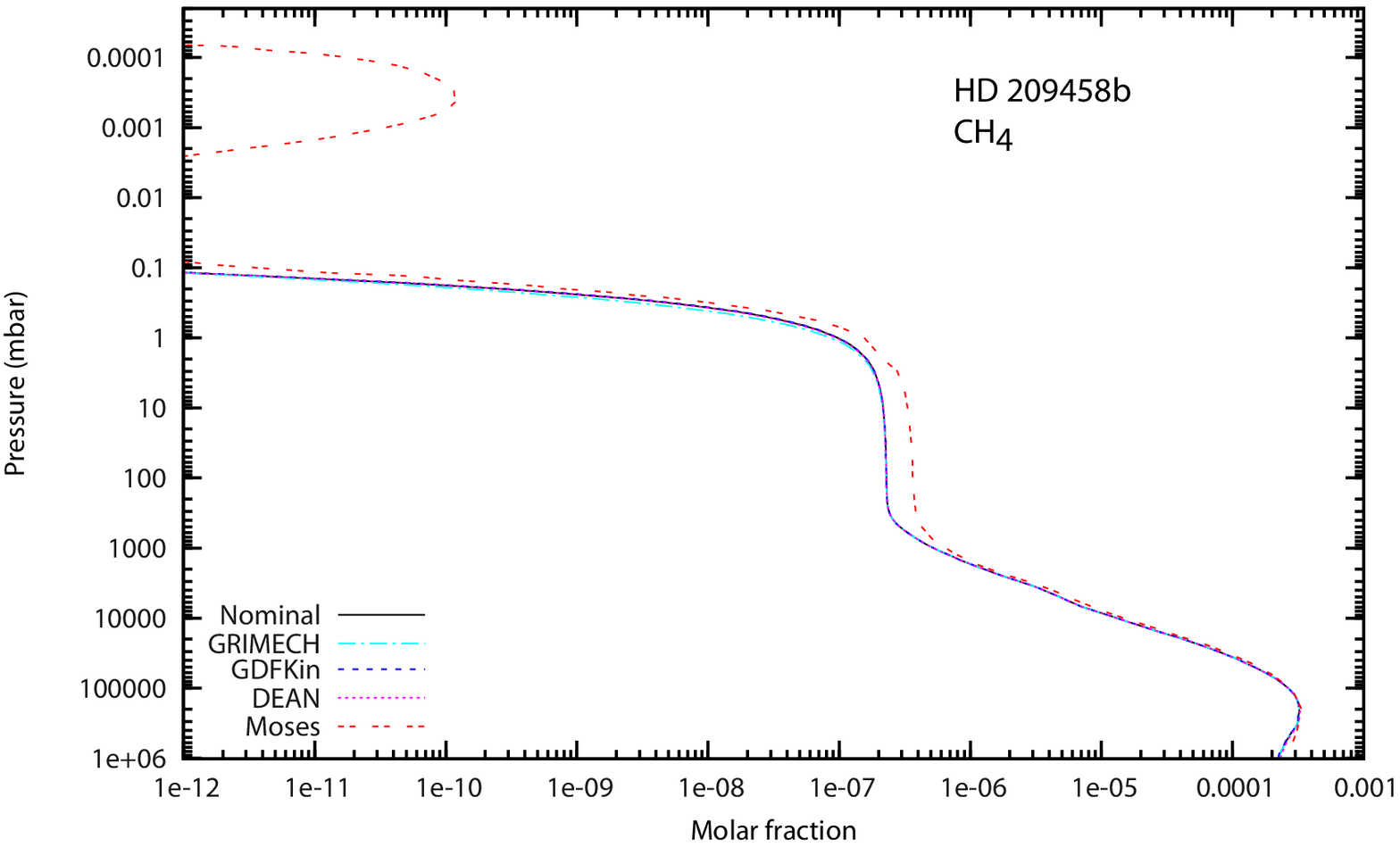}
\includegraphics[width=\columnwidth]{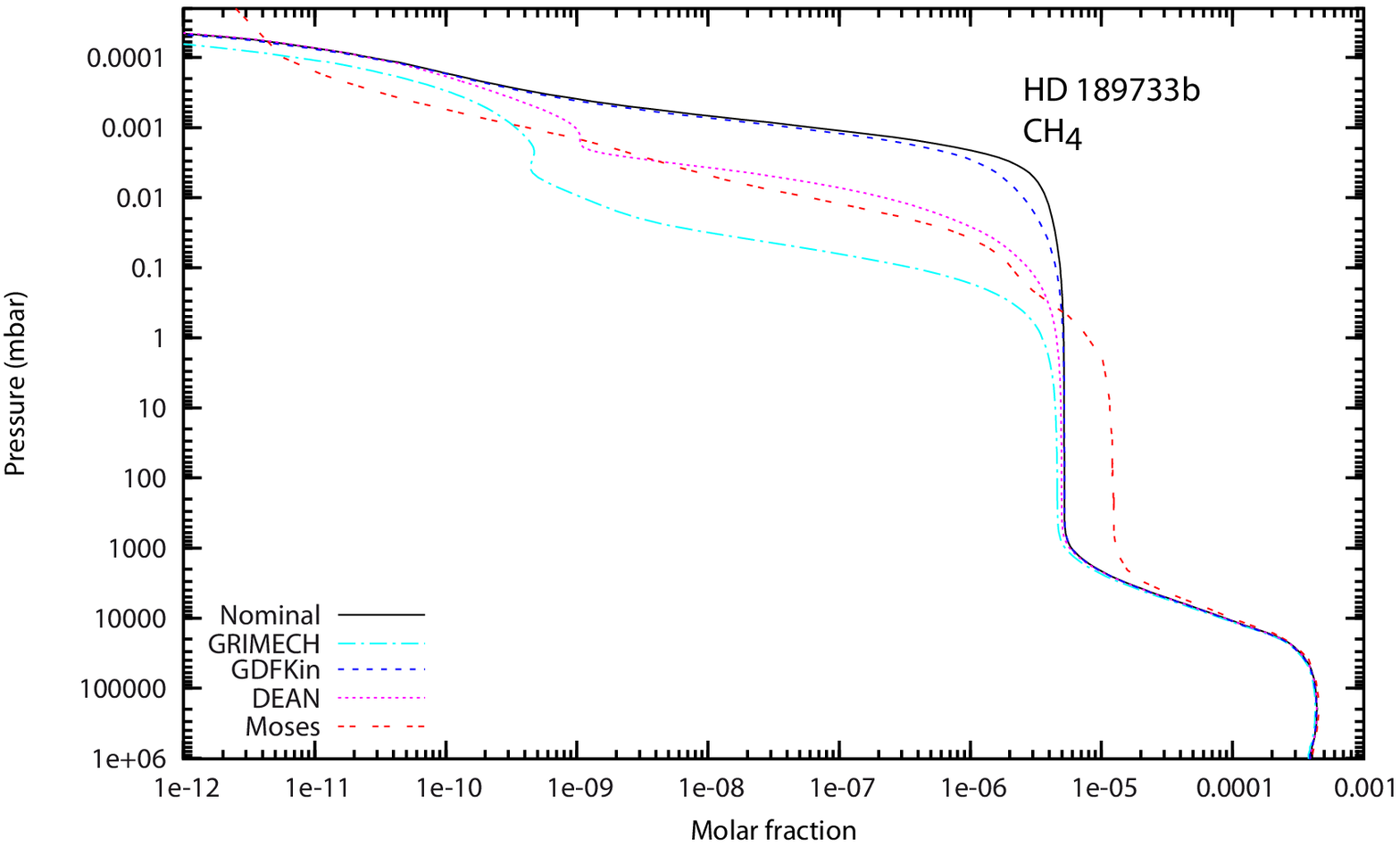}
\includegraphics[width=\columnwidth]{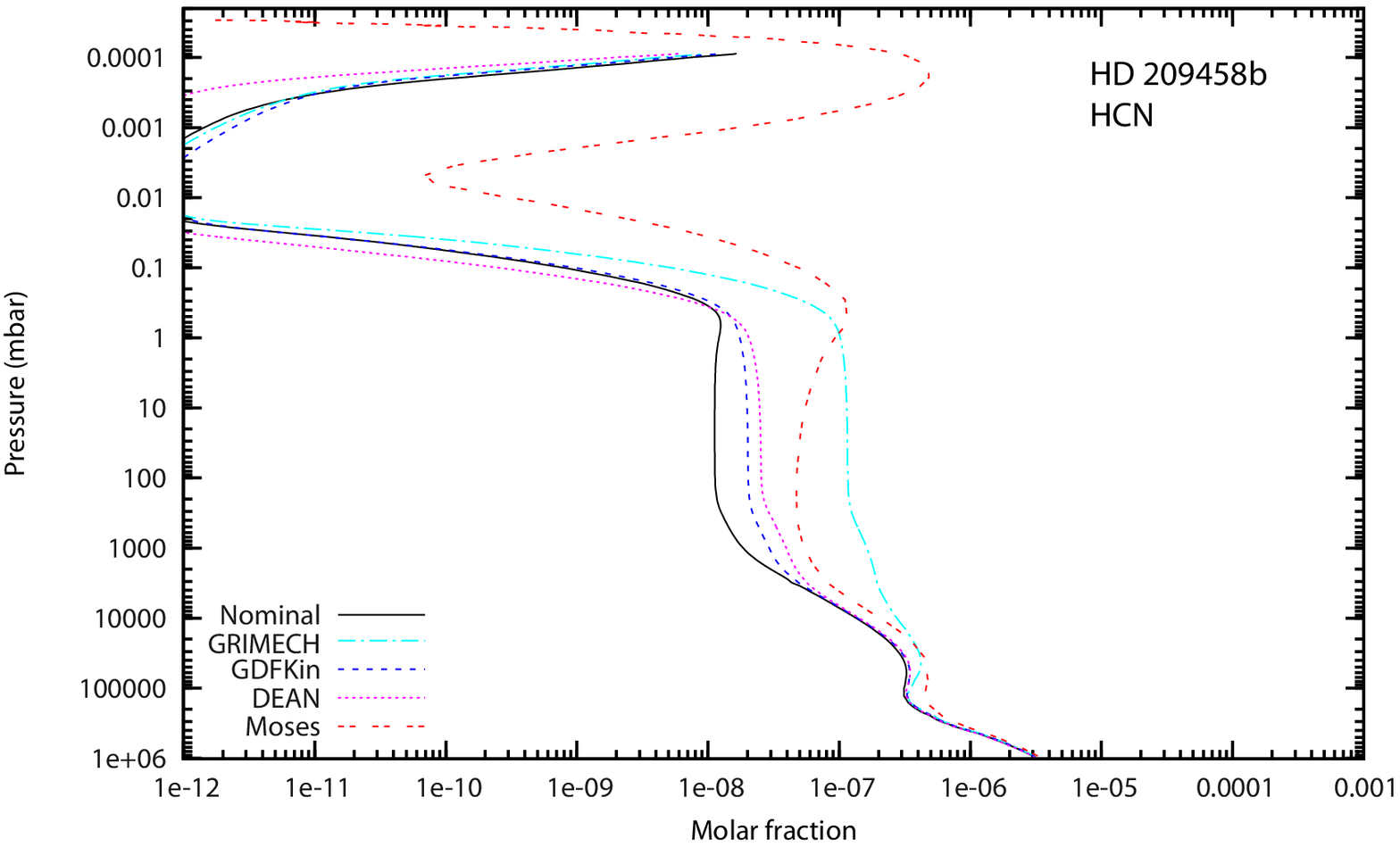}
\includegraphics[width=\columnwidth]{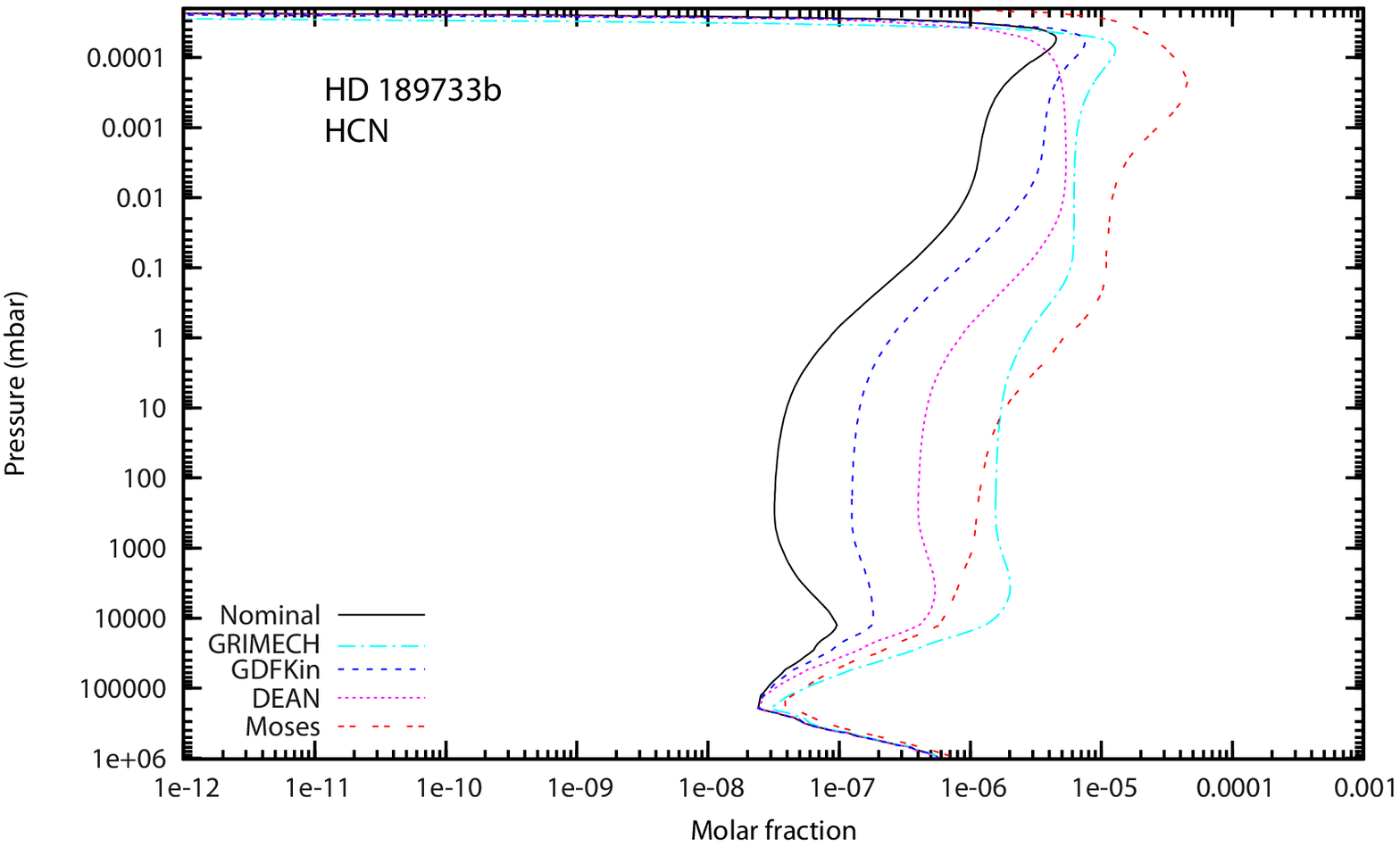}
\includegraphics[width=\columnwidth]{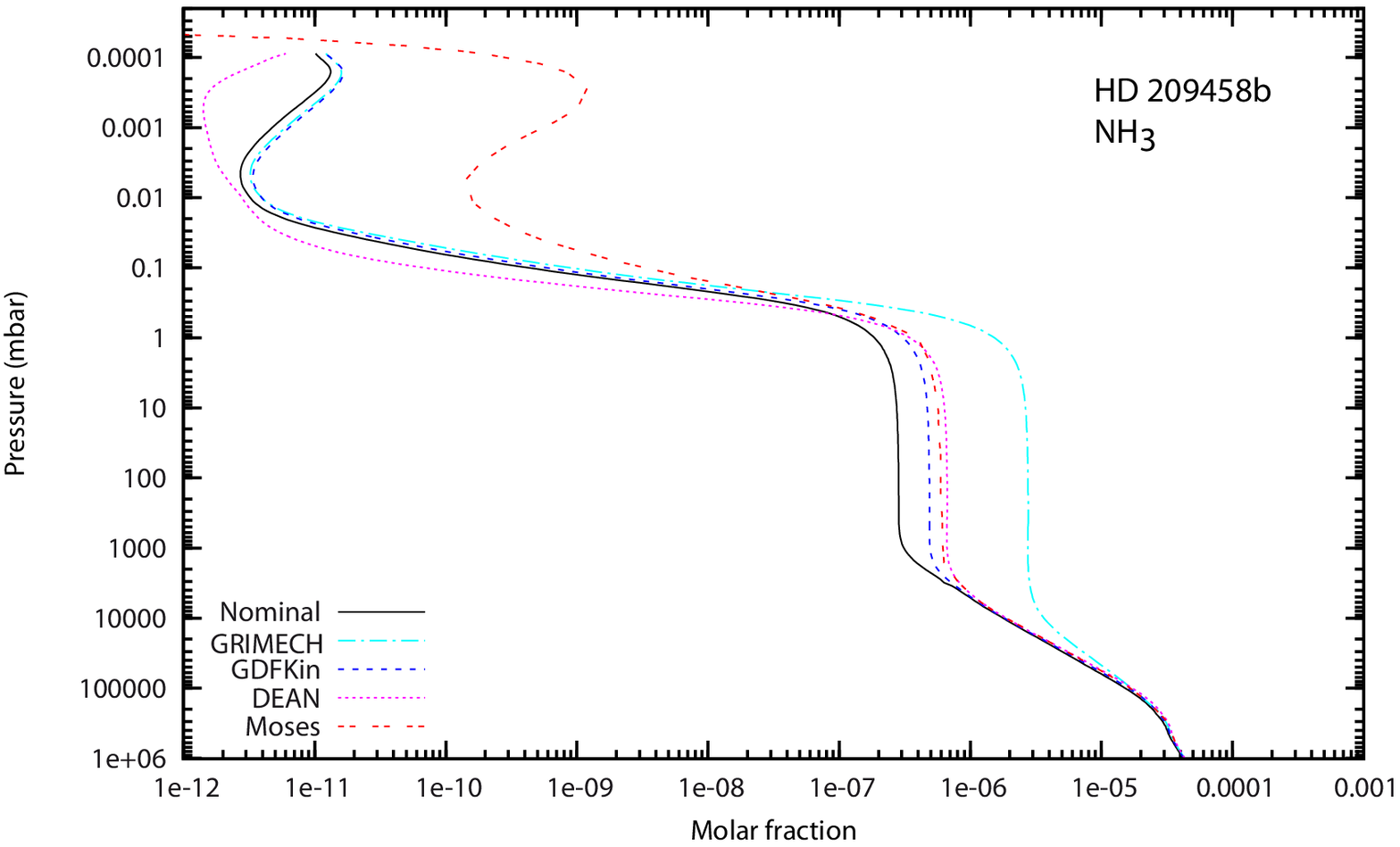}
\includegraphics[width=\columnwidth]{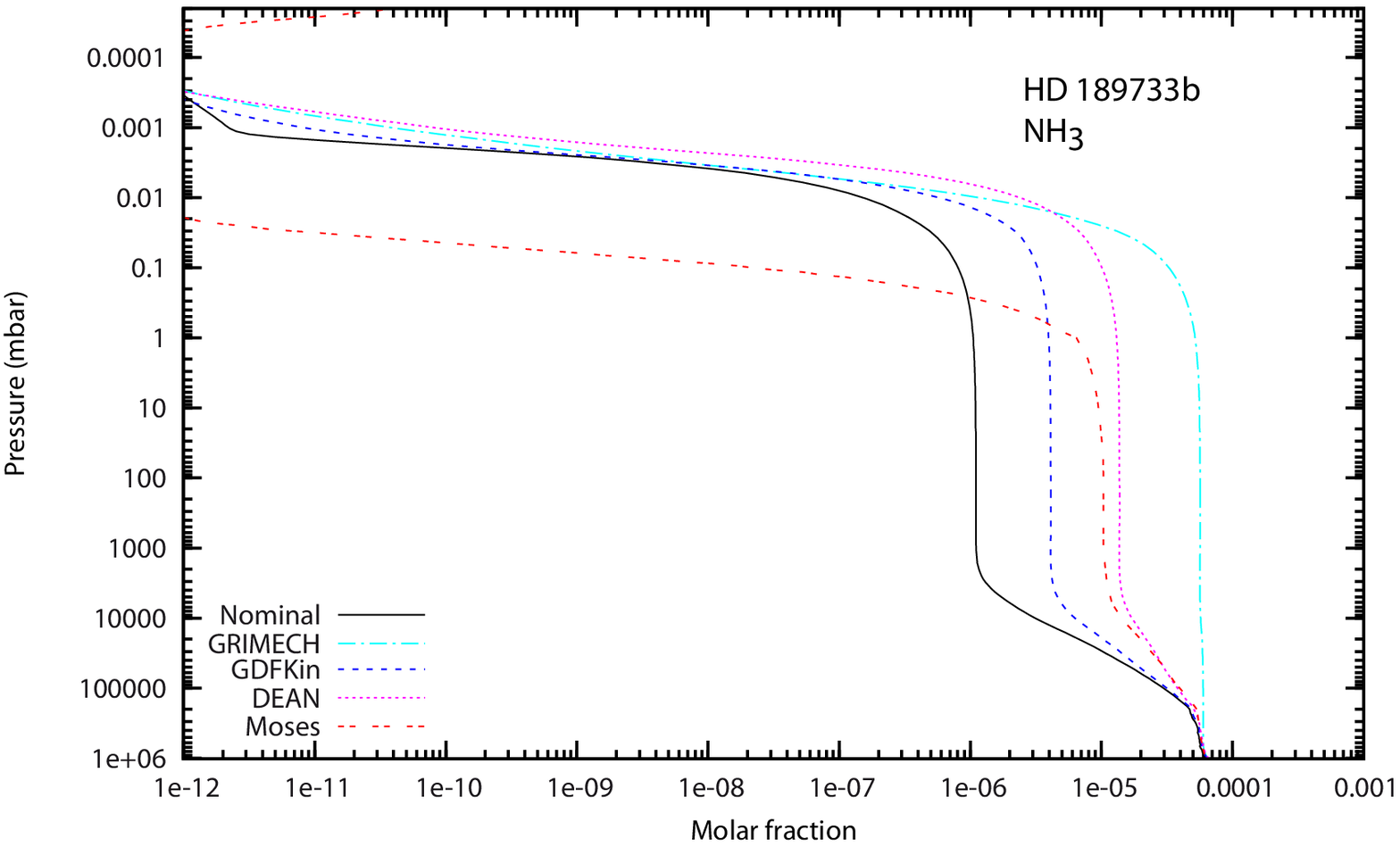}
\includegraphics[width=\columnwidth]{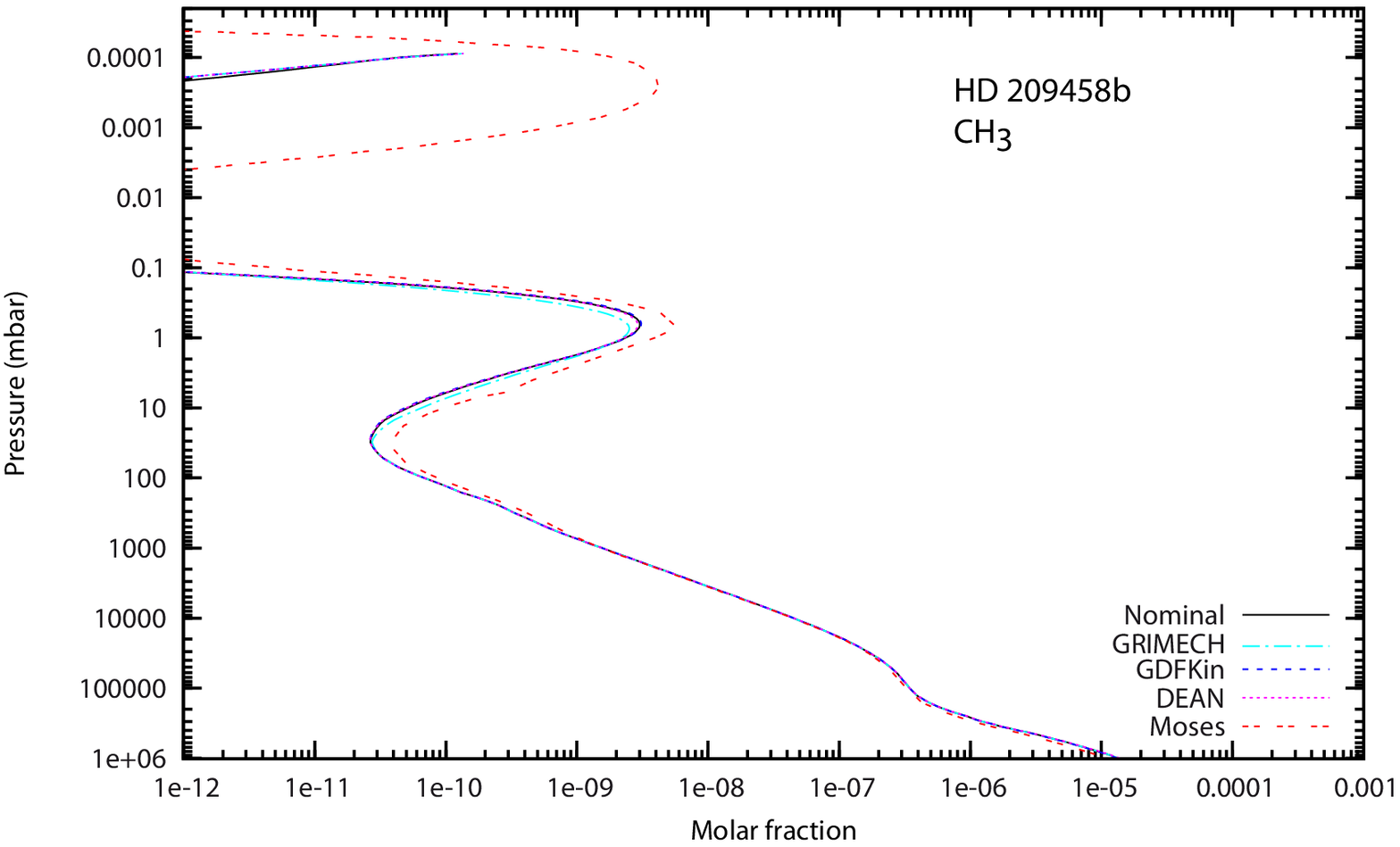}
\includegraphics[width=\columnwidth]{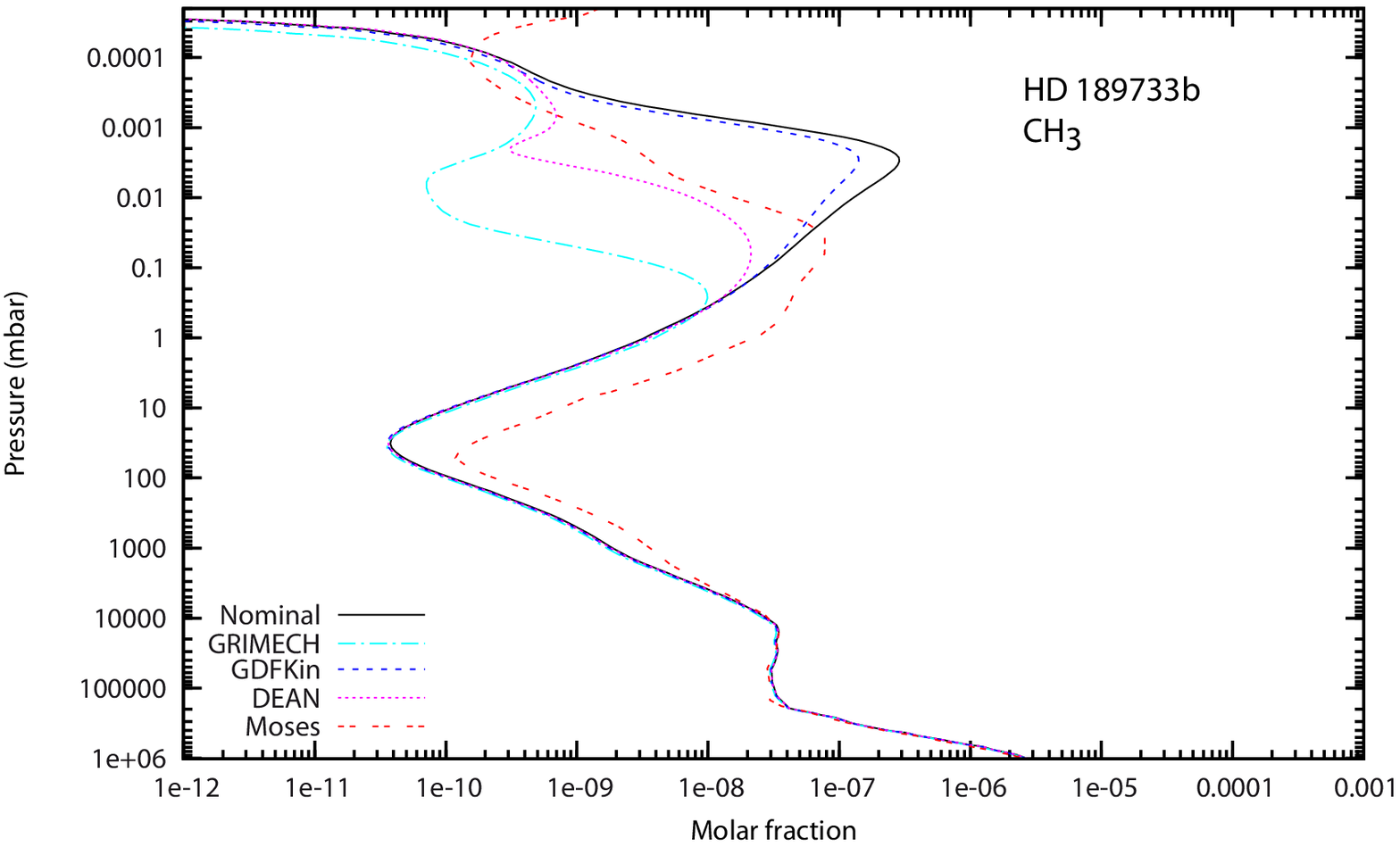}
\caption{Abundances of \ce{CH4}, HCN, \ce{NH3} and \ce{CH3} in HD~209458b (left) and HD~189733b (right) with the four different models, compared to the results of \citet{moses2011disequilibrium}.}
\label{fig:models1}
\end{figure*}

\begin{figure*}[!htbp]
\centering
\includegraphics[width=\columnwidth]{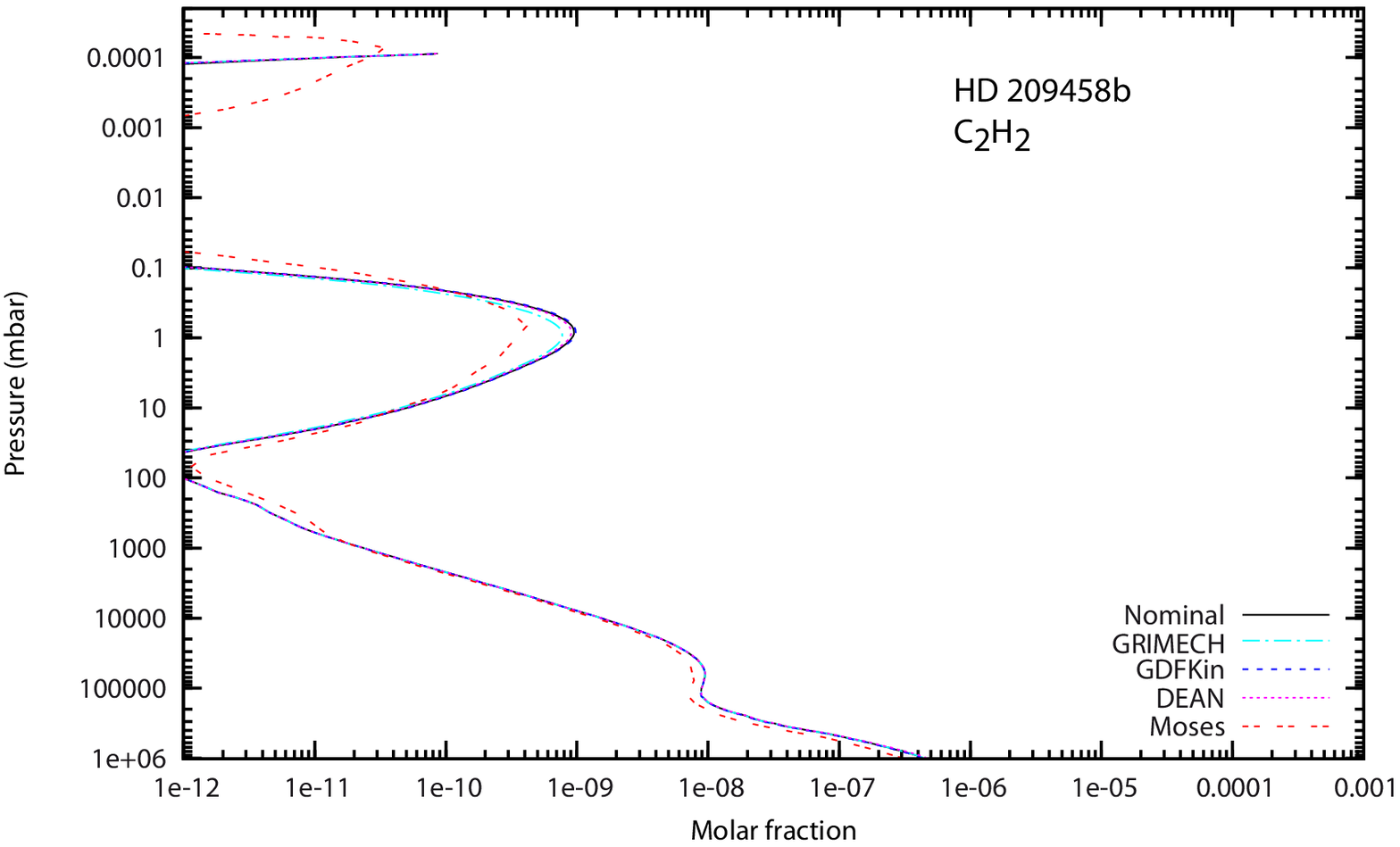}
\includegraphics[width=\columnwidth]{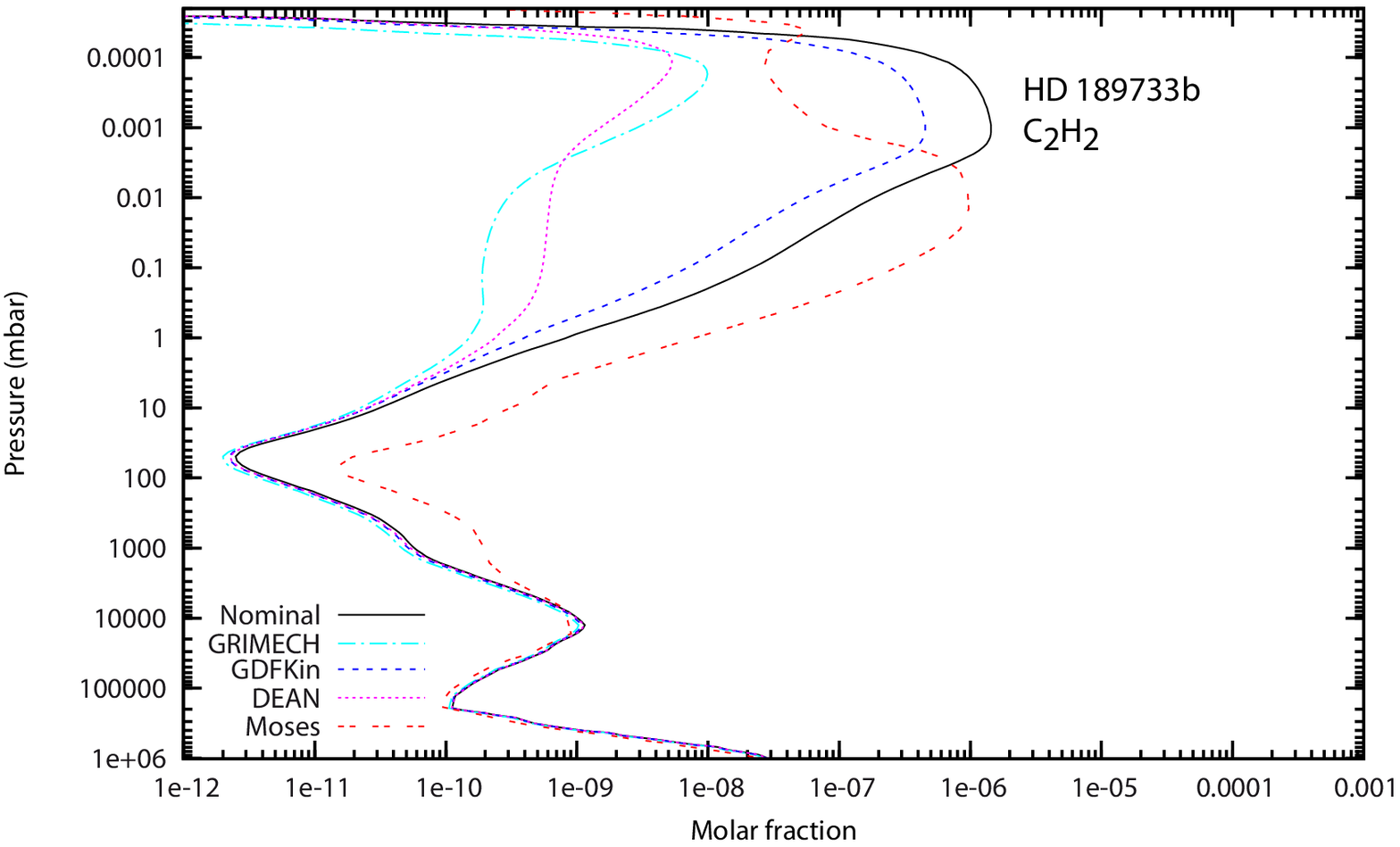}
\includegraphics[width=\columnwidth]{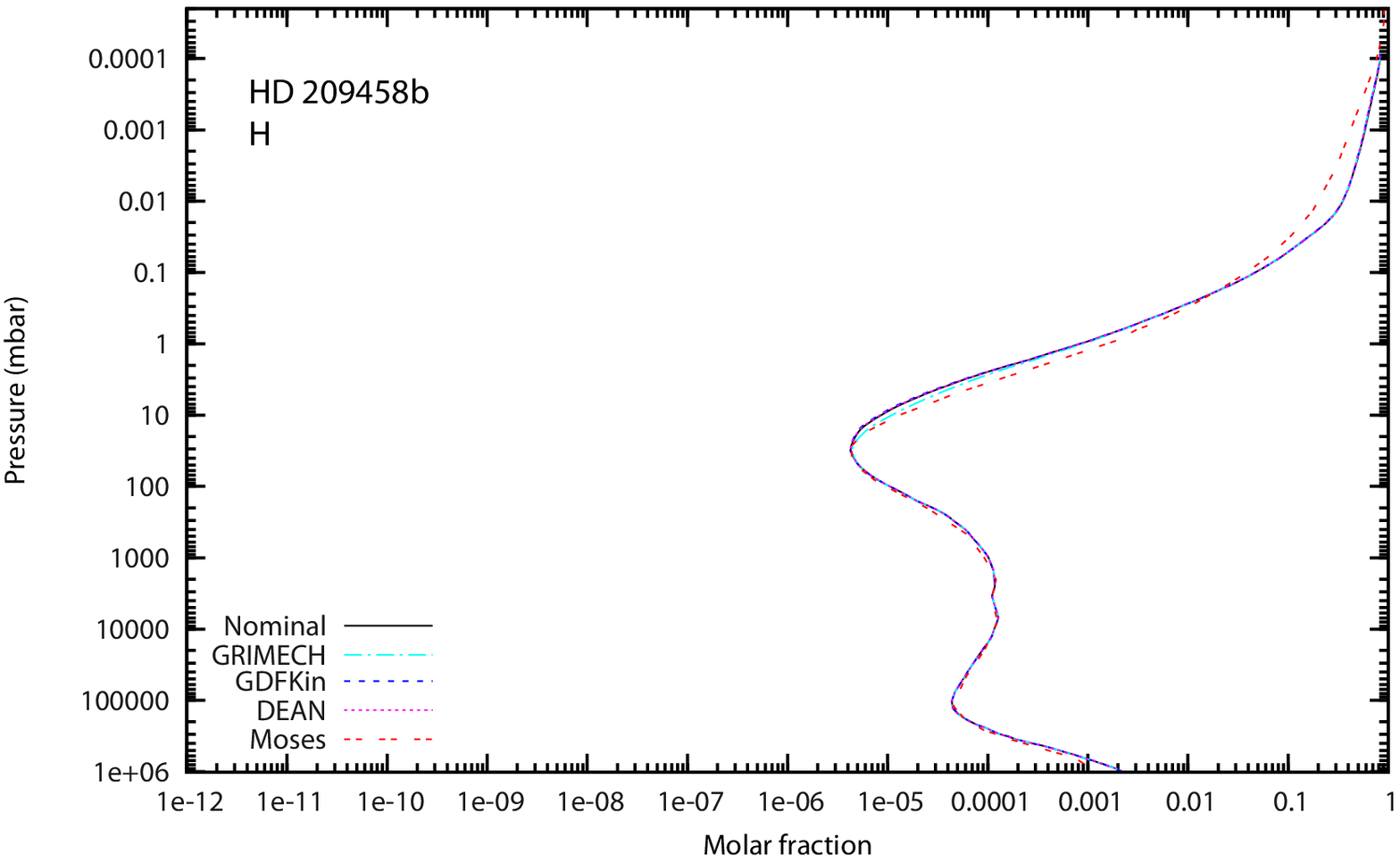}
\includegraphics[width=\columnwidth]{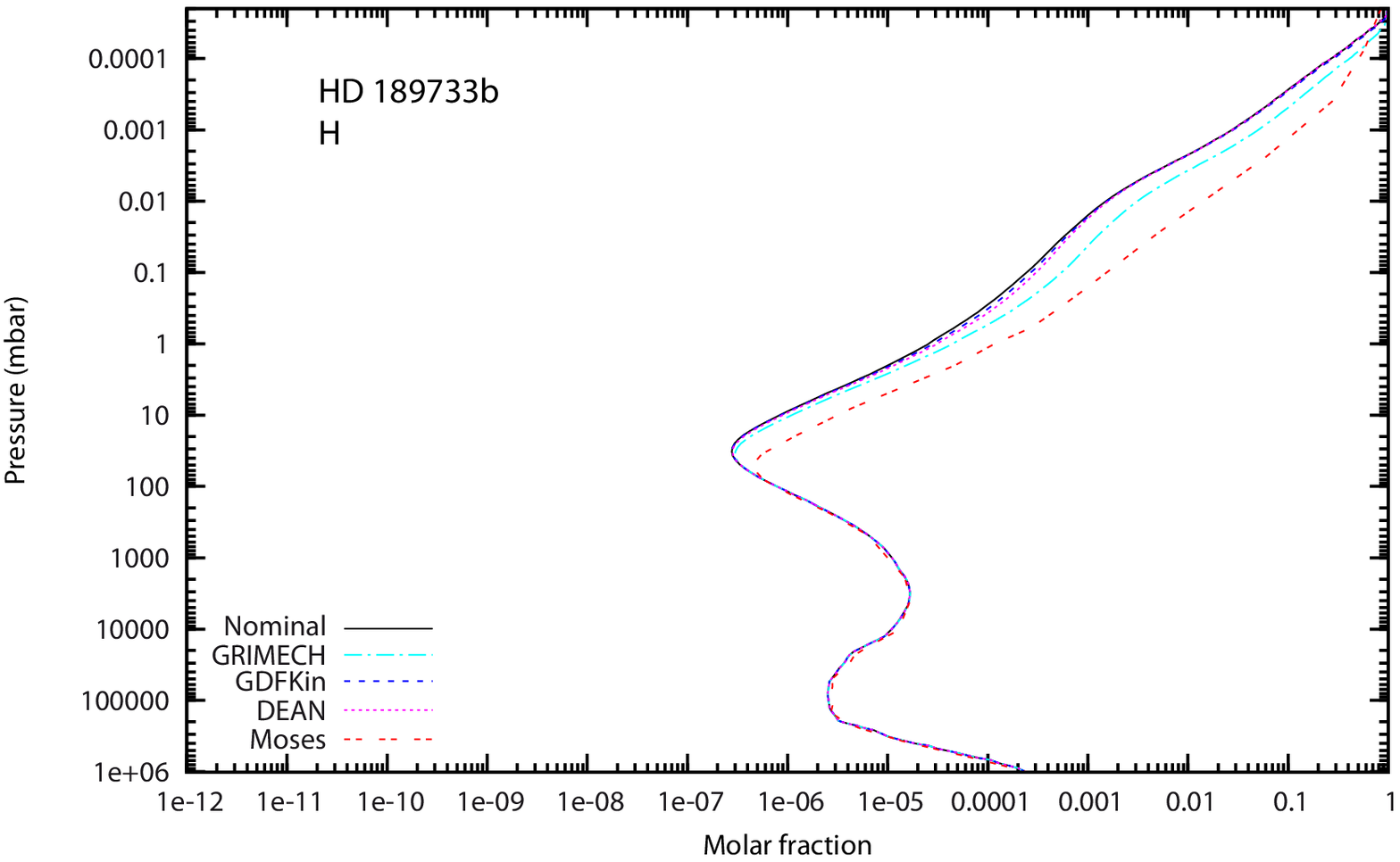}
\includegraphics[width=\columnwidth]{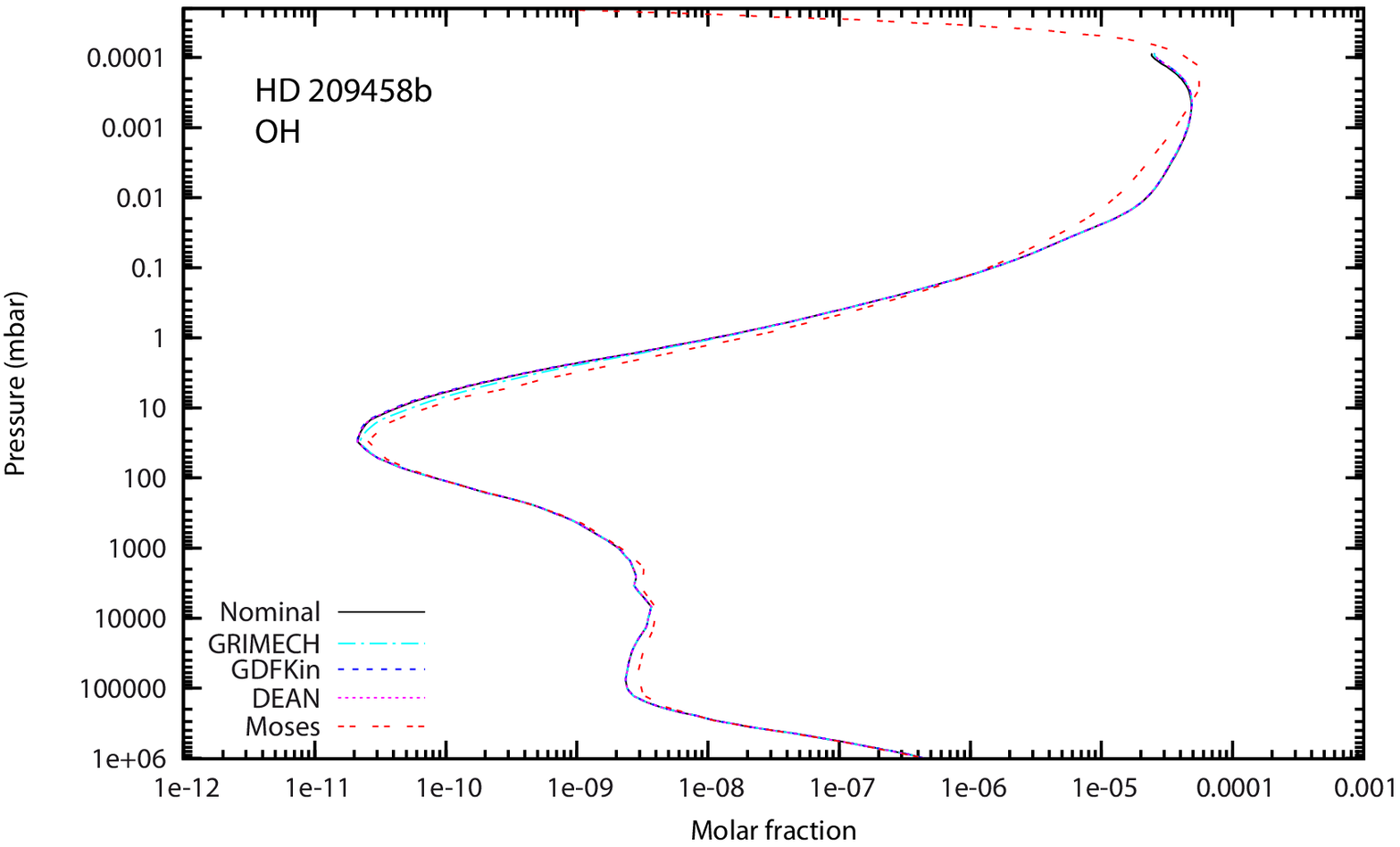}
\includegraphics[width=\columnwidth]{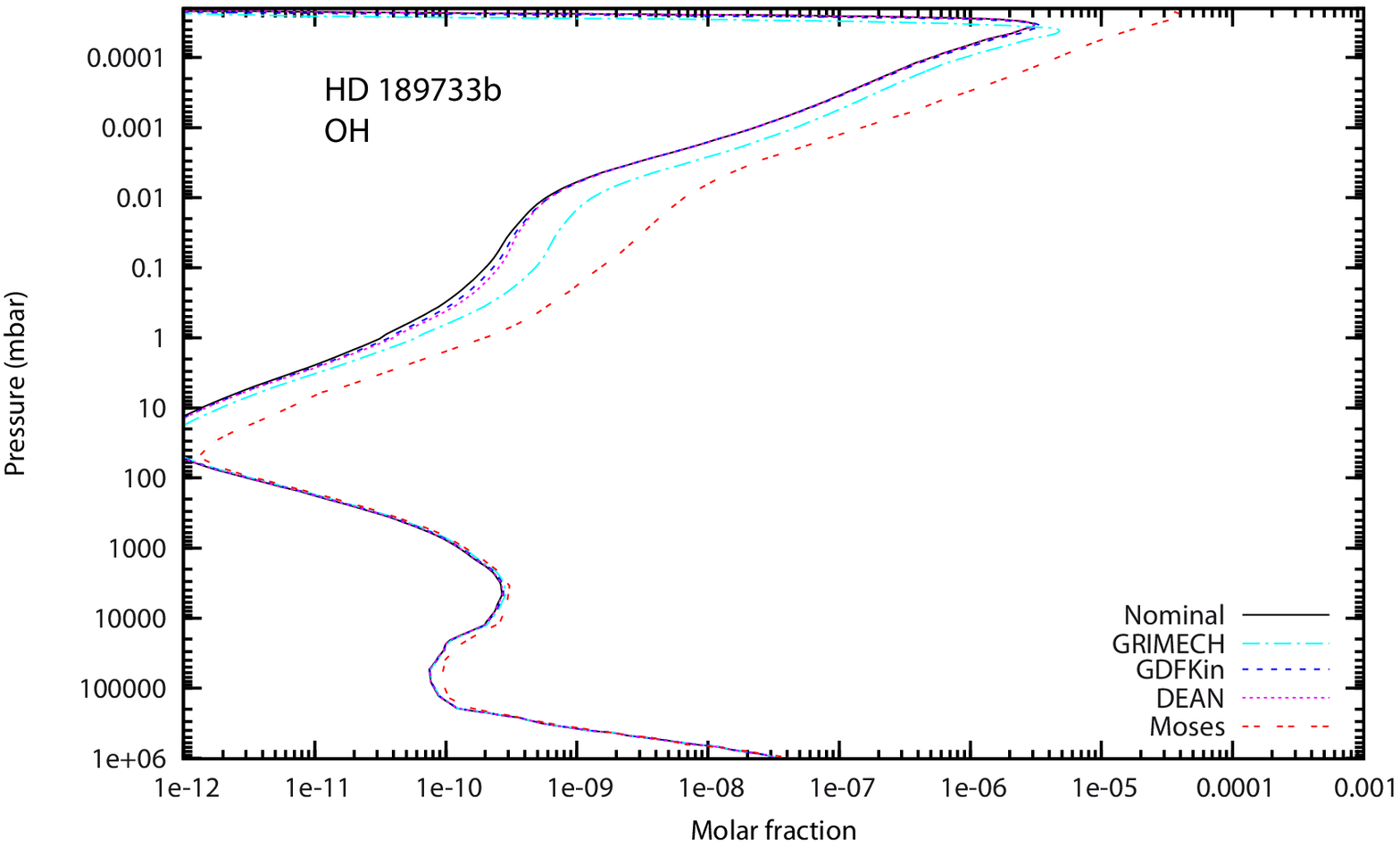}
\includegraphics[width=\columnwidth]{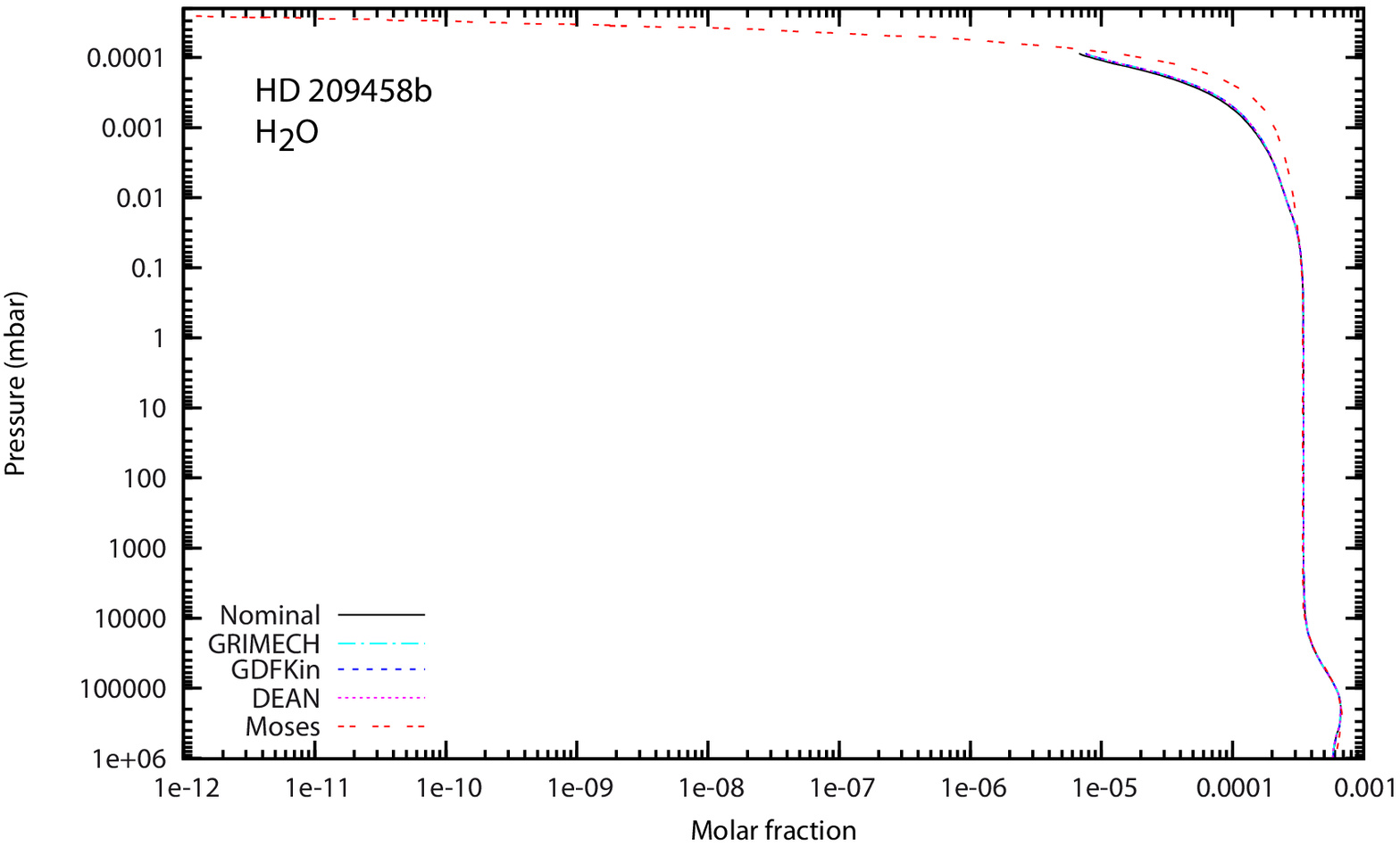}
\includegraphics[width=\columnwidth]{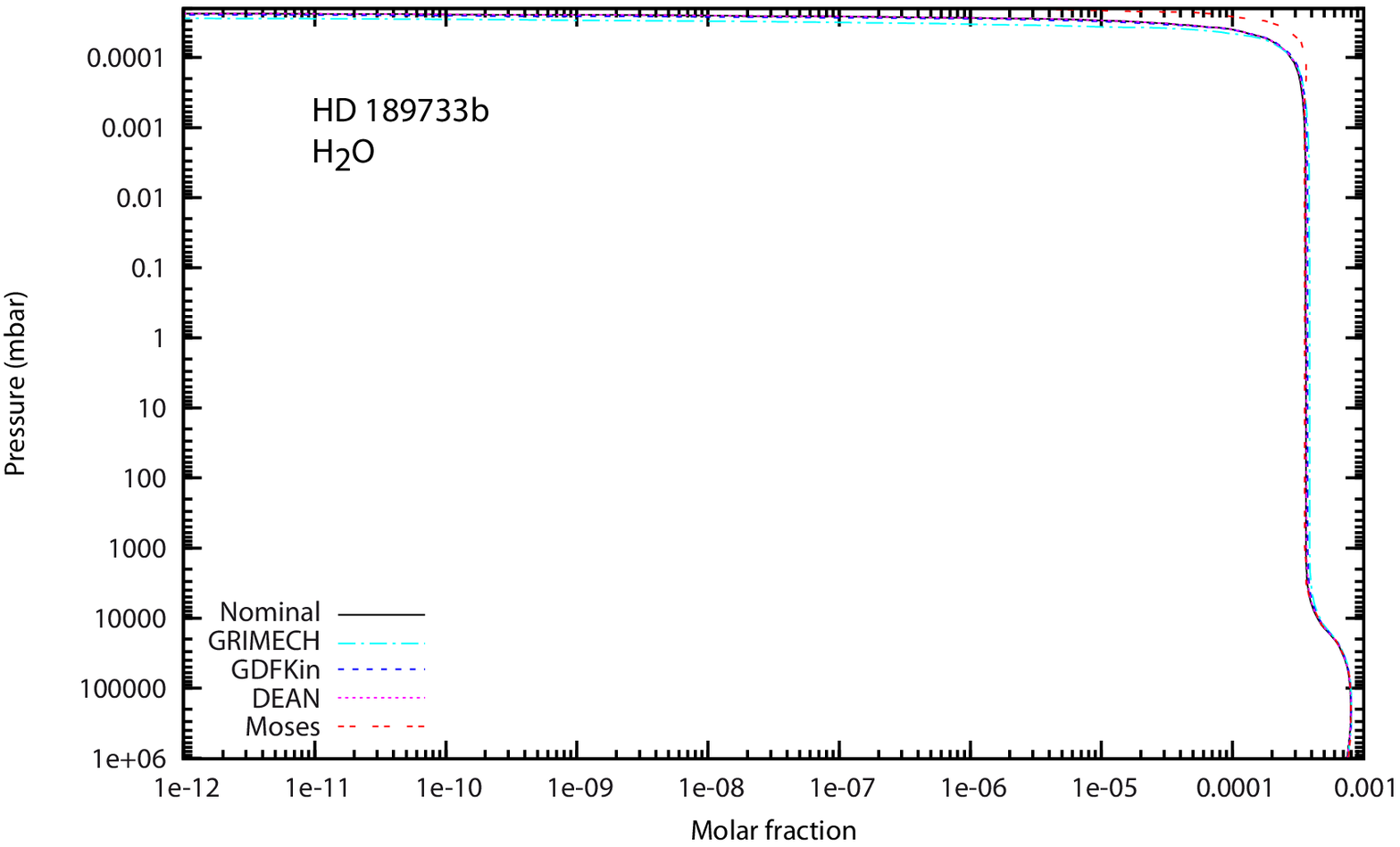}
\caption{Abundances of \ce{C2H2}, H, OH and \ce{H2O} in HD~209458b (left) and HD~189733b (right) with the four different models, compared to the results of \citet{moses2011disequilibrium}.}
\label{fig:models2}
\end{figure*}

The impact of the different nitrogen sub-mechanisms on the abundance profiles of various species is illustrated in Figs.~\ref{fig:models1} and \ref{fig:models2}. Thermodynamic equilibrium is the same for all schemes (for the species in common). 

First, we restrict our analysis to pressures higher than $\sim1$~mbar in order not to mix effects due to kinetic rates with possible differences in the photochemical data or modeling. For HD~209458b, the main species that are significantly affected at these pressure levels by the change of nitrogen scheme are HCN and \ce{NH3}. This is not surprising as these are the most abundant nitrogen species departing from equilibrium due to quenching, and the pressure level at which the quenching occurs depends on the kinetic network adopted. These two species (but also others) show even larger differences in the case of HD~189733b, as the lower temperatures of this atmosphere enhance the differences due to the kinetics. Departures between schemes are expected to become even more important for cooler atmospheres.
None of the tested schemes shows an general improvement of the agreement with M11. Similar \ce{NH3} quenching is found by M11 and with \textit{DEAN} for both planets, which makes sense as M11 use \citet{dean2000gas} as a source of reactions and associated rates for N-bearing compounds. This similarity is found also for HCN but only at pressures higher than 1~bar.
At higher altitudes, the HCN profiles from M11 become closer to the result with \textit{GRIMECH}. For both planets and both HCN and \ce{NH3}, profiles obtained with \textit{GDF-Kin} are bracketed by those from the nominal model and \textit{DEAN}, while \textit{GRIMECH} gives significantly higher abundances than all other models in the quenching region. With \textit{GRIMECH}, we also notice that \ce{NH3} becomes the main nitrogen-bearing species from the bottom of the atmosphere up to 0.03 mbar because of vertical mixing, whereas thermodynamics predicts that \ce{N2} should be the main nitrogen-bearing species.

Understanding the roots of theses discrepancies would require an in-depth study of the sensitivities of these schemes to reaction cycles, as a function of temperature and pressure, which is far beyond the scope of this study and would require tools that may have to be developed specifically for such large networks. To illustrate this difficulty let us identify the reaction that dominates the production rate of \ce{NH3} for HD~189733b, at 100 mbar. We find it to be the same for all mechanisms: \\
\begin{align}\label{reac:NH2+H2}
\ce{NH2 + H2 -> NH3 + H},
\end{align}
whose rate constant is similar in all schemes, and is calculated reversing the reaction:
\begin{align}\label{reac:NH3+H}
\ce{NH3 + H -> NH2 + H2},
\end{align}
which dominates the destruction of \ce{NH3}, also in all the schemes.
The following rates are found in the different schemes:\\
- Nominal, \textit{Dean, GRIMECH}: $9.00\times 10^{-19}T^{2.4}e^{-4990/T}$ cm$^3$.molecule$^{-1}$.s$^{-1}$, derived from \citet{ko1990rate}.\\
- \textit{GDF-Kin}: $1.056\times 10^{-18}T^{2.39}e^{-5114/T}$ cm$^3$.molecule$^{-1}$.s$^{-1}$ from \citet{michael1986rate}.\\
We could think that these slight differences are responsible for the different results. However, \ce{NH3} does not display the same abundance when using \textit{Dean} and \textit{GRIMECH} mechanisms, despite the fact that they both share the same rate constant. Moreover, nulling the rate constant of this reaction in the nominal scheme, does not affect the quenching level of \ce{NH3}, nor its abundance for pressures higher than 10 mbar. We can therefore eliminate this hypothesis. Key reactions are in fact usually those that limit the rate of a cycle and that do not dominate the production or destruction of a given species. Finding those limiting processes in complex networks is a field of research in itself. 
%- For \textit{GDF-Kin} and \textit{GRIMECH}, it is 
%\begin{align}
%\ce{CN + H2 -> HCN + H}
%\end{align}
%with $k_f=4.9\times10^{-19}T^{2.45}e^{-1130/T}$ cm$^3$.molecule$^{-1}$.s$^{-1}$ \citep{wooldridge1996shock}.\\
%- For the nominal one, it is
%\begin{align}
%\ce{HCNH -> HCN + H}
%\end{align}
%with $k_f=6.1\times10^{28}T^{-5.69}e^{-12,200/T}$ s$^{-1}$ \citep{dean2000gas}.\\
%- For \textit{Dean}, it is
%\begin{align}
%\ce{HNC -> HCN}
%\end{align}
%which is the reverse of the isomerization \ce{HCN -> HNC} with $k_f=2.6\times10^2T^{-3.23}e^{-24,950/T}$ s$^{-1}$ \citep{dean2000gas}. \\
Identifying key pathways and their limiting reactions require dedicated algorithms \citep{lehmann2004algorithm, grenfell2006chemical, Dobrijevic2010a, stock2011chemical, stock2012chemical} whose adaptation to the large networks we consider will require further work.

For hydrocarbons (see for instance \ce{CH4},  \ce{C2H2} and \ce{CH3}) all the models we tested cluster to the same profiles for pressures below 1~mbar. This shows that N-bearing species have little influence on hydrocarbon chemistry at these altitudes. (This would no longer be true at higher temperature and for higher C/O ratios as HCN would become a major reservoir of both N and C). M11 systematically finds higher mixing ratios for hydrocarbons (but within one order of magnitude) above the quenching level of \ce{CH4} (1-10~bar), likely due to kinetic differences in the C$_0$-C$_2$ mechanism.

At lower pressure, Figs.~\ref{fig:models1} and \ref{fig:models2} show large differences that are no longer due to quenching. At pressures lower than 1~mbar, the abundances of hydrocarbons depend on the nitrogen network used. It is particularly striking for \ce{C2H2} in HD~189733b, where \textit{DEAN} and \textit{GRIMECH}, on the one hand, and the nominal model, \textit{GDF-Kin} and M11, on the other hand, seem to cluster in two different regimes, exhibiting 2 to 3 order of magnitude differences at 0.1-0.001~mbar. Departures between network results can be due to differences in the kinetic network (different reactions, different rates, different minor species included) but also in photochemistry. Indeed, some UV-absorbing species are not included in all the models, such as \ce{N2H4}, \ce{HNO3}, \ce{C2N2} and \ce{N2O4}, which have absorption domains that overlap that of \ce{C2H2}.

\subsubsection{Corresponding emission and transmission spectra}\label{sec:spectra}

\begin{figure*}[!htbp]
\includegraphics[width=0.95\columnwidth]{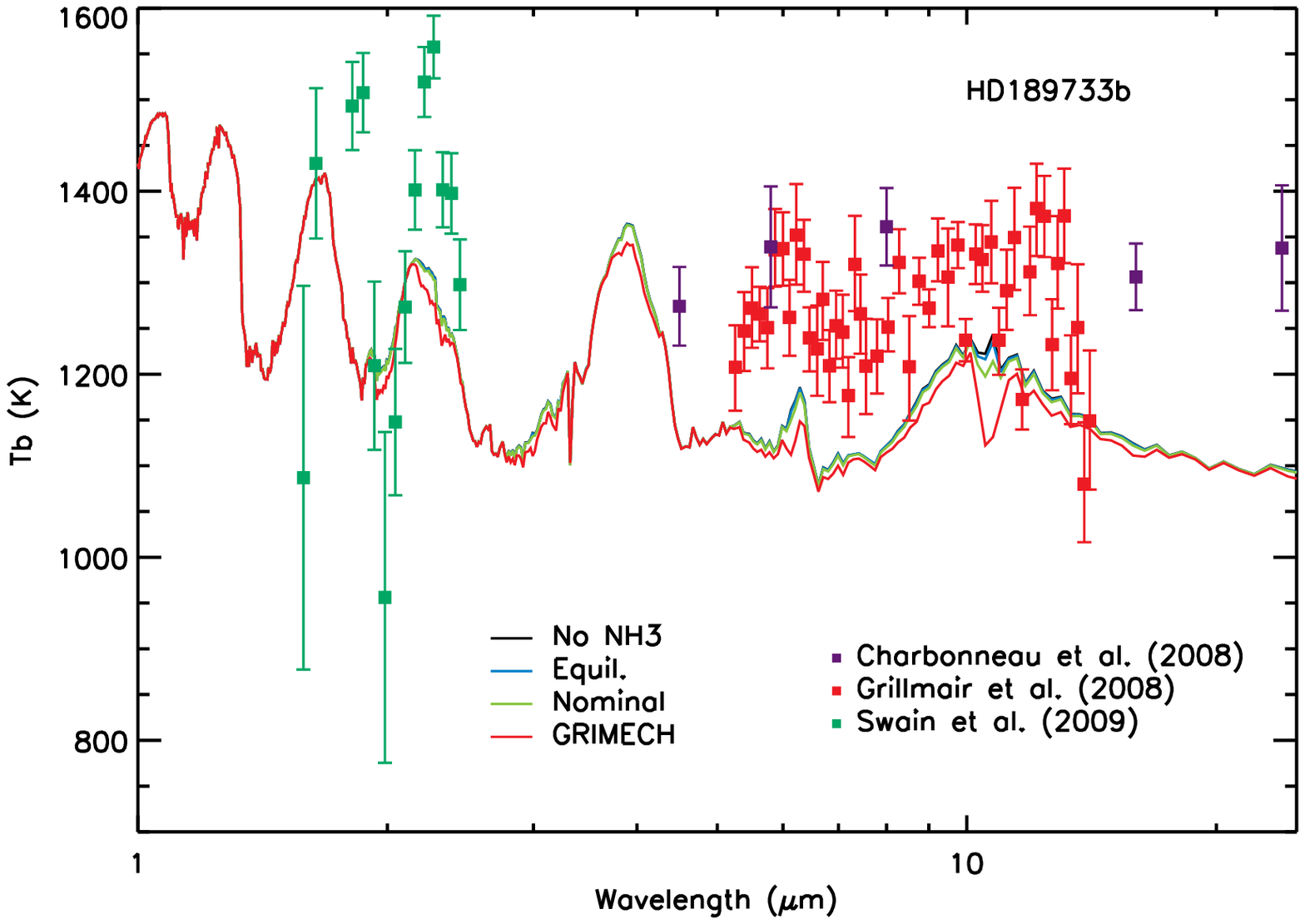}
\includegraphics[width=0.95\columnwidth]{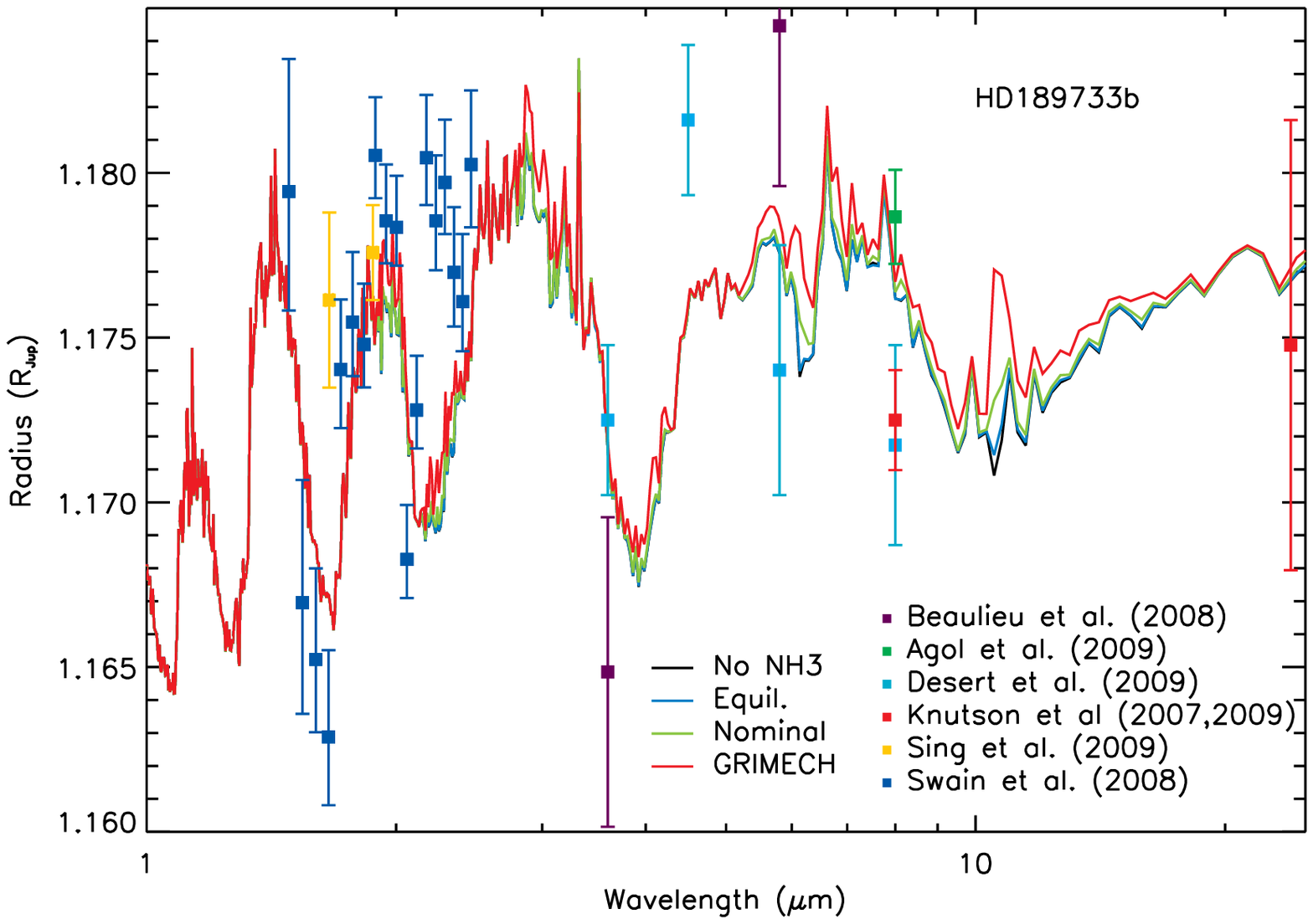}
\caption{Synthetic day-side (left) and transmission (right)  spectra of HD~189733b with the nominal mechanism (green curve) compared to the one corresponding to the \textit{GRIMECH} mechanism (red curve) and to the thermochemical equilibrium (blue curve). The dark curve is obtained when \ce{NH3} is removed from the model. The day-side fluxes are given as brightness temperatures (T$_b$).  Because of the reflection component, note that the link between T$_b$ and the atmospheric thermal profile is altered below 2 $\mu$m. The transmission spectrum is given as the apparent planetary radius. The data points obtained from various observations are also shown.}
\label{spectra}
\end{figure*}

In order to calculate the planetary transmission and emission+reflection spectra of HD~189733b (Fig.~\ref{spectra}) and HD~209458b, we use a line-by-line radiative transfer model from 0.3 to 25 $\mu$m \citep{Iro05,Iro10}. The opacity sources included in the model are the main molecular constituents: \ce{H2O}, CO, \ce{CH4}, \ce{NH3}, Na, K and TiO; Collision Induced Absorption by \ce{H2} and He; Rayleigh diffusion and \ce{H-} bound-free and \ce{H2-} free-free. For absorbing species not included in our kinetic model (Na, K and TiO), chemical equilibrium is assumed. The current model does not account for clouds. For the reflected component, we use synthetic stellar spectra generated from ATLAS\footnote{http://kurucz.harvard.edu/stars.html}. The main difference from the static model described in \cite{Iro05} is the addition of \ce{NH3} for which we used the HITRAN 2008 database \citep{HITRAN2008}. Planetary parameters are taken from Table~\ref{properties}.

We applied this model for the compositions obtained with the two nitrogen mechanisms, which give the most opposite results (Nominal and \textit{GRIMECH}), as well as for chemical equilibrium. The \textit{GRIMECH} scheme gives the highest abundance for ammonia: ten and one hundred times more \ce{NH3} than the nominal model for HD~209458b and HD~189733b, respectively. As a consequence, features of this molecule become noticeable on both the emission and transmission spectra at 1.9, 2.3, 3.0, 4.0, 6.1 and 10.5 $\mu$m. The most prominent feature is found for HD~189733b at 10.5 $\mu$m. \ce{NH3} features are also visible on the spectra of HD~209458b, but so slightly that it would not be observable.

At the moment, our radiative transfer model does not include the contribution of HCN to the opacities. Based on the HCN abundances and associated spectra found by M11, we can expect the spectra to be also sensitive to the HCN abundance. Indeed, at the altitudes probed by the observations, there is nearly two orders of magnitude less HCN with our nominal model than with \textit{GRIMECH}, and \textit{GRIMECH} gives HCN abundances similar to that of M11. Therefore, the signature of HCN found by M11 at 14~$\mu$m should also become noticeable with the \textit{GRIMECH} version of our scheme.

Some observational data are superimposed to the spectra \citep{charbonneau2008broadband, grillmair2008strong, swain2009molecular} (for emission spectrum) and \citep{knutson2007map, knutson2009multiwavelength, swain2008presence, beaulieu2008primary, sing2009transit, desert2009search, agol2010climate} (for transmission spectra), but we do not discuss the agreement between observations and synthetic spectra, as we did not attempt to fit the observable using different thermal profiles.

\section{Discussion}

%\subsection{Background atmosphere}

To study HD~209458b and HD~189733b, we use the pressure-temperature and eddy diffusion profiles from M11. These profiles are derived from general circulation models of \citet{showman2009atmospheric}. This choice is motivated by the comparison with M11 and also because the  actual physical structure of these atmospheres is not yet well constrained by observations. Note however, that \citet{2012arXiv1202.4721H} recently published thermal profiles for both planets, as inferred from transit spectroscopy. It seems that HD~189733b could be warmer than HD~209458b between $10^{-5}$ and $10^{-3}$ bar. Considering the large uncertainties affecting their physical conditions, our models should not be considered as predictions of the composition of hot Jupiters but more as a step in developing chemical models of these objects and model intercomparison.\\
In addition, circulation is not included in our model nor, to our knowledge, in other current photochemical models of hot Jupiters although it has a significant influence on the chemical composition of the atmospheres due, for instance, to the strong longitude dependency of the temperature. We study the influence of the horizontal transport on the composition of the atmosphere of HD~209458b in a forthcoming paper (Ag\'{u}ndez et al., in preparation).

%\subsection{nitrogen mechanism}
Because modeled abundances depend significantly on the reaction network (in particular for \ce{NH3}, HCN and some hydrocarbons), we recommend using a network that has been validated (but not optimized) against experiments for conditions as close as possible to those of application. We do not claim that the network we release is a definitive one, it will necessarily evolve as new experimental results and kinetic/thermodynamic data become available. Detailed nitrogen chemistry, for instance, has been implemented in combustion networks only recently and will be subjected to further evolution. Missing elements should also be added to the network, sulfur being the most obvious one. Since the scheme has not been optimized, adding new species and reactions to the network we release is possible. Moreover, although the range of \textit{P}, \textit{T} and the elements considered in combustion model do fit well with the study of hot Jupiter atmospheres, the ratio between hydrogen and the other elements does not. For this reason, among others, it is important to avoid using optimized networks that would prevent modeling of such hydrogen-rich mixtures. Note also that our current network, that cannot be used to study the abundance of species with more than 2 carbon atoms, is likely insufficient to study atmospheres with C/O ratios close to or above unity. For this reason we are currently working on an extended network that can model species up to 6 carbon atoms.\\
Hot Jupiter atmospheres represent an extreme case of planetary atmospheres in terms of both high temperatures and low metallicity. With the progress of observations, cooler and heavier atmospheres are or will be soon (with JWST, EChO, Finesse, E-ELT) accessible to characterization. Cooler atmospheres will depart more from equilibrium and will thus be more sensitive to the details of the kinetics. Molecules that remain minor constituents of hot Jupiter atmospheres will become more abundant in cooler atmospheres and have an increased influence on their spectral appearance and thermal structure, making the use of validated schemes even more relevant.  More metallic atmospheres should be found as we explore the exoplanet realm towards smaller objects with higher core/envelope masses, and eventually terrestrial objects. While some uncertainties still exist when applying our network to hydrogen-rich atmospheres, atmospheres with decreasing hydrogen to heavy elements ratio become closer to the conditions of validation (by their equivalence ratios, to use combustion terms). Even hot Jupiters can be significantly enriched (by factor of 10 or more) in heavy elements compared to their parent stars, which could for instance explain the high observed abundances of \ce{CO2} \citep{zahnle2009atmospheric}.

\section{Conclusion}

We have constructed a chemical scheme to study the atmospheric composition of hot Jupiters. Compared to existing models we chose to use one that \\
- is derived from mechanisms that are intensively used for industrial applications (in particular car engine simulations), \\
- has been subjected to in-depth validation protocols in a broad range of temperatures, pressures and compositions, \\
- is based on individual rate coefficient that have not been altered in order to optimize the agreement between the collective behavior of the network and experiments (contrary to most mechanisms in combustion). This allows users to apply the network slightly beyond its validation domain and to add species and reactions,\\
- uses experimental measurements for some of the endothermic reactions when robust data are available but still reproduces thermodynamic equilibrium with an excellent accuracy.

We developed a 1D photochemical model based on this kinetic scheme, and which includes vertical transport (mixing and molecular diffusion) and photodissociations. We applied this model to the hot Jupiters HD~209458b and HD~189733b and compared our results with those of \citet{moses2011disequilibrium}. Qualitatively, we find similar conclusions: photodissociations do not have a significant impact on the atmospheric composition of HD~209458b, with the high temperatures we assume. It remains at the thermodynamic equilibrium for pressures higher than 1 bar. For lower pressures vertical transport affects the abundances of HCN, \ce{NH3} and \ce{CH4} and some of the minor species associated. For HD~189733b,  we assume significantly lower temperatures and find the atmospheric composition to be more sensitive to photolyses and vertical transport, all species being affected, except the main reservoirs, \ce{H2}, \ce{H2O}, CO and \ce{N2}.
Quantitatively, however, we find significant differences (up to several orders of magnitude in the case of HD~189733b) in the abundances that are likely to be due to the different chemical schemes used. These differences are smaller for HD~209548b because kinetics have less influence. The quenching of HCN and \ce{NH3}, as well as \ce{CH4} to a lower extent, is particularly affected, as well as most species sensitive to photochemistry in the upper atmosphere. Despite being large in terms of abundances, these differences do not produce strong effects on the spectra, with the exception of \ce{NH3} at 10.5~$\mu$m. Confronting different schemes with observations will thus have to wait more accurate spectroscopic observations (JWST, E-ELT, EChO). Until that, experimental validation appears mandatory.\\
In order to illustrate the sensitivity to the kinetic scheme, we implemented different available nitrogen schemes, that are either optimized (\textit{GRIMECH}, \textit{GDFKin}) or non validated (\textit{Dean}). We studied the extent of the possible results, and found large differences whenever disequilibrium chemistry is at work. Changing the nitrogen scheme strongly affects the quenched species (HCN, \ce{NH3}) and most species (including hydrocarbons) in the upper atmosphere of HD~189733b. For HD~209458b, deviations are again less noticeable as the atmosphere departs less from equilibrium.  We therefore emphasize on the need to use validated and non optimized chemical schemes. This is already true for hot Jupiters but this is even more crucial in the case of cooler atmospheres (GJ1614b, GJ3470b, for instance), which depart more from thermodynamic equilibrium and are more sensitive to kinetics.\\
Our nominal scheme can be downloaded from the KIDA database\footnote{http://kida.obs.u-bordeaux1.fr/} \citep{wakelam_kida2012}. The scheme is designed to reproduce the kinetic evolution of species with less than two carbon atoms. In order to study atmospheres with C/O ratio higher than solar (close to or above 1), we are currently developing a C$_0$-C$_6$ scheme, which will be able to describe kinetics of species up to 6 carbon atoms. One of the next improvement of our model should be the addition of sulfur. As kinetics of nitrogen species is an active field of research, we expect regular updates of the network (which would be notified and available on KIDA).

Note also that conclusions of this study on the chemical composition of hot Jupiters, which derive from models using an average 1D vertical profile, will probably have to be revisited with the effects of atmospheric circulation.

\begin{acknowledgements} 
We thank Ignasi Ribas for providing us the UV flux of $\epsilon$~Eridani, proxy of HD~189733 and Anthony M. Dean for providing us the list of rate coefficients from his book in an electronic form. F. S., O. V., E. H. and M. A. acknowledge support from the European Research Council (ERC Grant 209622: E$_3$ARTHs).
\end{acknowledgements}

%\bibliographystyle{aa}
%\bibliography{bib_article}

\appendix
\section{Thermochemical data}\label{Appendix:A}

Thermochemical properties, such as enthalpies of formation, entropies and heat capacities are of great importance to ensure the consistency between the rate parameters of the forward and reverse elementary reactions. They are also useful to estimate the heat release rate. Thermochemical data for all molecules or radicals have been estimated and stored as 14 NASA polynomial coefficients, according to the \citet{mcbride1993nasa} formalism. The NASA polynomials have the following form:

\begin{align}\label{c_nasa}
\frac{c^0_p(T)}{RT} &= a_{1} + a_2T + a_3 T^2  + a_4 T^3+a_5 T^4\\
\label{h_nasa}
\frac{h^0(T)}{RT} &= a_1 + a_2 \frac{T}{2} +a_3 \frac{T^2}{3}  + a_4 \frac{T^3}{4}+a_{5}\frac{T^4}{5} +\frac{a_6}{T}\\
\label{s_nasa}
\frac{s^0(T)}{R} &= a_1 \ln T+ a_2  T + a_3\frac{T^2}{2}  + a_4 \frac{T^3}{ 3}+a_5\frac{T^4}{4} + a_{7}
\end{align}

Where $a_i$, $i \in [1,7]$, are the numerical NASA coefficients for the fourth-order polynomial. Each species is characterized by fourteen numbers. The first seven numbers  are for the high-temperature range, generally from 1000 to 5000 K, and the following seven numbers are the coefficients for the low-temperature range, generally from 300 to 1000~K. When these parameters are not available in the literature \citep{mcbride1993nasa, allendorf, goose2010} which is the most frequent case for species present in automotive fuels, they have to be estimated. In this case, these data were automatically calculated using the software THERGAS \citep{muller1995thergas}, developed in the LRGP laboratory, based on the group and bond additivity methods proposed by \citet{Benson76} and updated based on the data of \citet{domalski}. The enthalpies of formation of alkyl radicals have been also updated according to the values of bond dissociation energies published by \citet{tsang1986chemical} and by \citet{luo2003handbook} and following the recommendations of \citet{Benson97} respectively.

An elementary reversible reaction {\it i} involving {\it L} chemical species can be represented in the following general form:
\begin{equation}
\sum_{l=1}^{L}\nu_{li}'\chi_l \Leftrightarrow \sum_{l=1}^{L}\nu_{li}''\chi_l
\end{equation}

\noindent where $\nu_{li}'$ are the forward stoichiometric coefficients, and $\nu_{li}'' $ are the reverse ones. $\chi_l$ is the chemical symbol of the $l^{th}$ species.

The kinetic data associated to each reaction are expressed with a modified Arrhenius law $k(T) = A\times T^n \exp^{-\frac{E_a}{RT}}$ where {\it T} is the temperature, $E_a$ is the activation energy of the reaction, {\it A} the pre-exponential factor and {\it n} a coefficient which allows the temperature dependence of the pre-exponential factor.
If the rate constant associated to the forward reaction is $k_{fi}(T)$, then the one associated to the reverse reaction is $k_{ri}(T)$, verifying:

\begin{equation}
K_{pi} = \frac{k_{fi}(T)}{k_{ri}(T)} \left( \frac{k_BT}{P^0}\right)^{\sum_{l=1}^L \nu_{li}}
\end{equation}
where $K_{pi}$ is the equilibrium constant, when the activity of the reactants is expressed in pressure units \citep{Benson76}:
\begin{equation}
K_{pi} = \exp\left({\frac{\Delta S_i^0}{R} -\frac{\Delta H_i^0}{RT}}\right)
\end{equation}

\noindent where $\Delta S_i^0$ and $\Delta H_i^0$ are the variation of entropy and enthalpy occurring when passing from reactants to products in the reaction~{\it i}, $P^0$ is the standard pressure ($P^0$ =1,01325 bar), $k_B$ is the Boltzmann's constant and $\nu_l$ are the stoichiometric coefficients of the $L$ species involved in reaction {\it i}: $\nu_l=\nu_{li}''-\nu_{li}'$. 
Combined with Eqs. (\ref{h_nasa}) and (\ref{s_nasa}), $\frac{\Delta S_i^0}{R}$ and $\frac{\Delta H_i^0}{RT}$ can be calculated with the NASA coefficients:

\begin{equation}
\frac{\Delta S_i^0}{R} = \sum_{l=1}^L \nu_{l}\frac{s^0_l(T)}{R}  \qquad \mathrm{and} \qquad
\frac{\Delta H_i^0}{RT} = \sum_{l=1}^L \nu_{l}\frac{h^0_l(T)}{RT}
\end{equation}
Finally, we can calculate the reverse reaction rate for the reaction $i$:
\begin{equation}
k_{ri}(T) = \frac{k_{fi}(T)}{K_{pi}}\left( \frac{k_BT}{P^0}\right)^{\sum_{l=1}^L \nu_{li}}
\end{equation}

\section{Chemical equilibrium calculation} \label{Appendix:TECA}

To compute the equilibrium abundance of the species in a definite system considered as an ideal gas, we have developed a Thermodynamical Equilibrium Calculator TECA. TECA is a software which allows the equilibrium calculation for a complex mixture. More specifically, for a given initial state of an ideal-gas mixture, the chemical-equilibrium program is able to determine the gas composition at a defined temperature and pressure. This calculation is based on the principle of the minimization of Gibbs energy \citep[e.g.][]{gibbs1873method, white1958chemical, eriksson1971thermodynamic, smith1982chemical, reynolds1986element}:

\begin{equation}
G = \sum_{l=1}^{L} \overline{g_l} N_l
\end{equation}

\noindent where $L$ is the total number of species, $\overline{g_l}$ is the partial free energy of the species $l$ and $N_l$ is the number of moles of the species $l$.

The partial free energy of a compound $l$, behaving as an ideal gas, is given by:

\begin{equation}
\overline{g_l} = g_l(T,P) + RT \ln N_l
\end{equation}

where $g_l(T,P)$ is the free energy of the species $l$ at the temperature $T$ and the pressure $P$ of the system and $R$ is the ideal gas constant.

For an ideal gas, $g_l(T,P)$ is given by:

\begin{equation}
g_l(T,P) = h^0_l(T) - Ts^0_l(T) + RT \ln\left(\frac{P}{P^0}\right)
\end{equation}
where  $h^0_l(T)$ and $s^0_l(T)$ are respectively, the standard-state enthalpy and entropy of the species $l$ at the temperature $T$ of the system.

The enthalpy and the entropy are expressed as NASA polynomials as described above.

\section{Pressure-dependent reactions}

Under some conditions, several reactions do not have the same rate constant depending if they occur under low or high pressure (respectively $k_0(T)$ and $k_{\infty}(T)$). In this case, between these two limits it appears what is called a fall-off zone. This is typically the case in reactions requiring a collisional body to proceed, such as thermal dissociation or recombination (three-body) reactions.
In the present kinetic model we have different types of reactions with pressure dependent rate constants (Table~\ref{tab:threebody}). 
In some cases, some species act more efficiently as collisional bodies than do others. Then, when available from literature, collisional efficiencies are used to specify the increased efficiency of the $l^{th}$ species in the $i^{th}$ reaction (see for example reaction (2) in Table~\ref{tab:threebody}).

For the pressure-dependent reactions, the rate constant at any pressure is taken to be:

\begin{equation}
k(T)=k_\infty(T) \left( \frac{P_r}{1+P_r}\right)F
\end{equation}
 
\noindent where the reduced pressure $P_r$ is given by:

\begin{equation}
P_r=\frac{[M]k_0}{k_\infty}
\end{equation}
 
\noindent and $[M]$ is the concentration of the mixture, weighted by the efficiency of each compound, $\alpha_l$, in the reaction studied:

\begin{equation}
[M] = \sum_{l=1}^L \alpha_l [X_l]
\end{equation}

\noindent where $[X_l]$ is the concentration of the species k.

As shown in Table~\ref{tab:threebody}, three methods of representation of the rate expression in the fall-off region are used (enhanced collisional body efficiencies of certain species are presented below the reaction):

\begin{itemize}
\item the \citet{lindemann1922discussion} formulation, illustrated by reaction (1) in Table \ref{tab:threebody};
\item the \citet{troe1983theory} formulation, see for example reaction (2) in Table \ref{tab:threebody};
\item the SRI formulation proposed by \citet{stewart1989pressure}, illustrated by reaction (3) in Table \ref{tab:threebody}.
\end{itemize}

\begin{table*}[!htb]
\begin{tabularx}{0.70\textwidth}{ c l c c c c c }

& & & &  $\tt{k=A T^n exp(-E/RT)}$ & \multicolumn{2}{c}{\tt{- High pressure limit}}\\
 & \tt{Reaction considered} & & & \tt{A (cm$^3$.molecule$^{-1}$.s$^{-1}$.K$^{-n}$)} & \tt{n}  &\tt{E/R (K)} \\

 \tt{1.} & \tt{C2H4+OH(+M)=C2H4OH(+M)} & & &$\tt{9.003 \times 10^{-12}}$ & \tt{0.0} & \tt{0.0}\\
  & \: \tt{Low pressure limit:} &$\tt{3.284\times10^{-21}}$ & \tt{-3.1} & \tt{0.0}&&\\
 \tt{2.} &\tt{H+CH3(+M)=CH4(+M)}& & &$\tt{2.774\times 10^{-10}}$& \tt{0.0} & \tt{0.0}\\
 & \: \tt{O2} &  \tt{Enhanced by } & \tt{0.40} & & & \\
 & \: \tt{CO} &  \tt{Enhanced by } & \tt{0.75} & & & \\ 
 & \: \tt{CO2} & \tt{Enhanced by } & \tt{1.50} & & & \\ 
 & \: \tt{H2O} &  \tt{Enhanced by } & \tt{6.50} & & & \\ 
 & \: \tt{CH4} &  \tt{Enhanced by } & \tt{0.00}\footnote{\ce{M=CH4} has specific coefficients} & & & \\ 
 & \: \tt{C2H6} & \tt{Enhanced by } & \tt{3.00} & & & \\ 
 & \: \tt{Ar} & \tt{Enhanced by } & \tt{0.35} & & & \\ 
 & \: \tt{N2} &  \tt{Enhanced by } & \tt{0.40} & & & \\ 
 & \: \tt{He} &  \tt{Enhanced by } & \tt{0.35} & & & \\ 
& \: \tt{Low pressure limit:} &$\tt{3.885\times10^{-24}}$ & \tt{-1.8} & \tt{0.0}&&\\
& \: \tt{TROE centering} &\tt{0.37} & \tt{3315} & \tt{61}&&\\

 \tt{3.} & \tt{NO+OH(+M)=HONO(+M)}& & &$\tt{1.827\times 10^{-10}}$& \tt{-0.3} & \tt{0.0}\\
 & \: \tt{Low pressure limit:} &$\tt{6.484\times10^{-25}}$ & \tt{-2.4} & \tt{0.0}&&\\
 & \: \tt{SRI centering:} &\tt{1.0} & \tt{0.0} & $\tt{1.00\times10^{-18}}$&\tt{0.81}&\tt{0.0}\\
\end{tabularx}
\caption{Some example of reactions with pressure-dependent rate constants present in the kinetic model.}\label{tab:threebody}
\end{table*}

\noindent In the Lindenman form, {\it F} is unity ({\it F}=1).\\
In the Troe form {\it F} is given by

\begin{equation}
\log_{10} F = \frac{\log_{10}(F_{cent})}{1+\left[ \frac{\log_{10}(P_r) + c}{N - d(\log_{10}(P_r) + c)}\right ]^2}
\end{equation} 

\noindent with

\begin{align*}
   c&=-0.4-0.67\times \log_{10}(F_{cent}) \\
   N&=0.75-1.27\times\log_{10}(F_{cent}) \\
   d&=0.14
\end{align*}

\noindent and 

\begin{equation}
F_{cent} = (1-a) \exp\left({-\frac{T}{T^{***}}}\right) + a \exp \left(-\frac{T}{T^*}\right) + \exp\left(-\frac{T^{**}}{T}\right)
\end{equation}

The four parameters {\it a}, {\it T***}, {\it T*} and {\it T**} must be specified but it is often the case that the parameter {\it T**} is not used by lack of data.

The approach taken at the Stanford Research Institute (SRI) by \citet{stewart1989pressure} is in many ways similar to that taken by Troe, but the blending function {\it F} is approximated differently. Here, {\it F} is given by

\begin{equation}
F=d \left[ a\exp\frac{-b}{T}+\exp\frac{-T}{c} \right]^X T^e
\end{equation}

\noindent where 

\begin{equation}
X=\frac{1}{1+(\log_{10} P_r)^2}
\end{equation}

\section{Photodissociations}\label{Appendix:photodisso}

\begin{landscape}
\null\vfill
\tiny
\begin{table}[!h]

\begin{tabularx}{\linewidth}{l l l l r  }

% \toprule \addlinespace[2pt]
 \multicolumn{3}{l}{Pathways} & Cross sections  & Quantum yields  \\
 \hline
 %endfirsthead
% \midrule
% \multicolumn{2}{l}{\scriptsize{Pathways}} & \scriptsize{Cross sections} & \scriptsize{F$_{\sigma}$} & \scriptsize{Quantum yields} &  & \scriptsize{F$_q$} \\
% \hhline{-------}
%\endhead
% \midrule
%\endfoot
% \bottomrule
%\endlastfoot

J1 & \ce{H2O} + $h\nu$ &\ce{-> H + OH} & \cite{Chan1993387} and & \cite{huebner1992solar} \\

J2 &  &\ce{-> H2 + O(^1D)} & \cite{fillion2004high} and &  \\

J3 &  &\ce{-> H + H + O(^3P)} & \cite{mota2005water} &  \\

J4 & \ce{CO2} + $h\nu$ &\ce{-> CO +  O(^3P)} & \cite{stark2007, ityakov2008}, and &   \cite{huebner1992solar}\\

J5 &  &\ce{-> CO +  O(^1D)} & \cite{huestis2010} &  \\

J6 & \ce{H2CO} + $h\nu$ &\ce{-> H2 + CO} & \cite{cooper1996absolute} &   \cite{huebner1992solar} \\

J7 &  &\ce{-> H + HCO} & and \cite{meller2000temperature} & \\

J8 & \ce{OH} + $h\nu$ &\ce{-> H + O(^1D)} & \cite{huebner1992solar} & \cite{van1984dissociation}\\

J9 & \ce{CO} + $h\nu$ &\ce{-> C + O(^3P)} & \cite{olney1997absolute} & \cite{huebner1992solar} \\

J10 & \ce{H2} + $h\nu$ &\ce{-> H + H} & \cite{samson1994total, chan1992absolute} & Estimation \\
 & & & \cite{olney1997absolute} & \\
J11 & \ce{CH4} + $h\nu$ &\ce{-> CH3 + H} & \cite{au1993valence} and & \cite{gans2011}\\

J12 & &\ce{-> ^1CH2 + H2} & \cite{lee2001enhancement} and &  \\

J13 &  &\ce{-> ^3CH2 + H + H} & \cite{kameta2002photoabsorption} and &  \\

J14 &  &\ce{-> CH + H2 + H} & \cite{chen2004temperature} & \\

J15 & \ce{CH3} + $h\nu$ &\ce{-> ^1CH2 + H} & \cite{khamaganov2007photolysis} & \cite{Parkes1973425} \\

J16 & \ce{C2H2} + $h\nu$ &\ce{-> C2H + H} & \cite{cooper1995absolute, wu2001measurements}  & \cite{okabe1981photochemistry, okabe1983}\\

J17 & \ce{C2H3} + $h\nu$ &\ce{-> C2H2 + H} & \cite{fahr1998ultraviolet} & \cite{fahr1998ultraviolet}\\

J18 & \ce{C2H4} + $h\nu$ &\ce{-> C2H2 + H2} & \cite{cooper1995absoluteethylene}  & \cite{holland1997photoabsorption} \\

J19 &  &\ce{-> C2H2 + H + H} & \cite{orkin1997rate} and \cite{wu2004measurements} & \cite{chang1998} \\

J20 & \ce{C2H6} + $h\nu$ &\ce{-> C2H4 + H2} & \cite{au1993valence, Kameta1996391} & \cite{akimoto1965}\\

J21 &  &\ce{-> C2H4 + H + H} & \cite{lee2001enhancement} & \cite{hampson1965vacuum}\\

J22 & &\ce{-> C2H2 + H2 + H2} & \cite{chen2004temperature} &  \cite{lias1970}     \\

J23 &  &\ce{-> CH4 + ^1CH2} & & \cite{mount1978photoabsorption}  \\

J24 & &\ce{-> CH3 + CH3} &        &  \cite{mount1978photoabsorption} \\

J25 & \ce{N2} + $h\nu$ &\ce{-> N(^2D) + N(^4S)} & \cite{samson1964absorption, huffman1969absorption}& Estimation\\
& & & \cite{stark92, chan1993absolute}  & \\

J26 & \ce{HCN} + $h\nu$ &\ce{-> CN + H} & Lee (1980)    & \cite{lee1980cn} \\

J27 & \ce{NH3} + $h\nu$ &\ce{-> NH2 + H} & \cite{burton1993electronic, chen1999, cheng2006absorption} &    \cite{mcnesby1962vacuum} \\

J28 & \ce{NO} + $h\nu$ &\ce{-> N(^4S) + O(^3P)} & \cite{iida1986absolute, chan1993absoluteno} &   \cite{huebner1992solar} \\

J29 & \ce{N2H4} + $h\nu$ &\ce{-> N2H3 + H} & \cite{vaghjiani1993} & \cite{vaghjiani1993, vaghjiani1995laser} \\

J30 & \ce{HNO3} + $h\nu$ &\ce{-> NO2 + OH} & \cite{sander2011} & Estimation \\

J31 & \ce{C2N2} + $h\nu$ &\ce{-> CN + CN} &  B\'{e}nilan et al. (2012), in preparation &  \cite{cody1977, jackson1979multiphoton, eng1996energy} \\

J32 & \ce{N2O4} + $h\nu$ &\ce{-> NO2 + NO2} & \cite{vandaele1998measurements, merienne1997no} & \cite{sander2011}\\

J33 & \ce{N2O3} + $h\nu$ &\ce{-> NO2 + NO} & \cite{stockwell1978near} &  \cite{sander2011}\\

J34 &\ce{HCO} + $h\nu$ &\ce{-> H + CO} & \cite{hochanadel1980ultraviolet, loison1996} & Estimation 

\end{tabularx}
\caption{Photodissociations scheme used in the model}
\end{table}
\label{tab:photodissociations}
\vfill\null
\end{landscape}

\end{document}